\documentclass[%
 reprint,
 amsmath,amssymb,
 aps,
]{revtex4-2}

\usepackage{graphicx}
\usepackage{xcolor}
\usepackage{amsmath}
\usepackage{caption}
\usepackage{subcaption}
\usepackage{dcolumn}
\usepackage{bm}
\usepackage{caption}
\usepackage{subcaption}
\usepackage{hyphenat}
\usepackage{url}
\usepackage{svg}

\begin{document}

\title{Oscillatory and Excitable Dynamics in an Opinion Model with Group Opinions}

\author{Corbit R. Sampson}
 \affiliation{Department of Applied Mathematics, University of Colorado at Boulder, Colorado 80309, USA}
 
 \author{Juan G. Restrepo}
\affiliation{Department of Applied Mathematics, University of Colorado at Boulder, Colorado 80309, USA}

 \author{Mason A. Porter}
 \affiliation{Department of Mathematics, University of California, Los Angeles, California 90095, USA; Department of Sociology, University of California, Los Angeles, California 90095, USA; and Santa Fe Institute, Santa Fe, New Mexico 87501, USA}

\date{\today}


\begin{abstract}

In traditional models of opinion dynamics, each agent in a network has an opinion and changes in opinions arise from pairwise (i.e., dyadic) interactions between agents. However, in many situations, groups of individuals {possess} a collective opinion that {can} differ from the opinions of {its constituent} individuals. In this paper, we study the effects of group opinions on opinion dynamics. We formulate a hypergraph model in which both individual agents and groups of 3 agents have opinions, and we examine how opinions evolve through both dyadic interactions and {group memberships.} In some parameter regimes, we find that the presence of group opinions can lead to oscillatory and excitable opinion dynamics. In the oscillatory regime, the mean opinion of the agents in a network has self-sustained oscillations. In the excitable regime, finite-size effects create large but short-lived opinion swings (as in social fads). We develop a mean-field approximation of our model and obtain good agreement with direct numerical simulations. We also show {---} both numerically and via our mean-field description {---} that oscillatory dynamics occur only when the number of dyadic and polyadic interactions per agent are not completely correlated. Our results illustrate how polyadic structures, such as groups of agents, can have important effects on collective opinion dynamics.
\end{abstract}

\maketitle

\section{Introduction} \label{sec:level1}

The opinions of individuals in a social network often change when they are exposed to the opinions and actions of other individuals. The ensuing opinion dynamics of such individuals has received considerable attention from sociologists \cite{kozitsin2023opinion}, economists \cite{zha2020opinion}, political scientists \cite{fernandez2014voter, siegel2009social}, applied mathematicians and theoretical physicists \cite{fernandez2014voter,horstmeyer2020adaptive,deffuant2000mixing,neuhauser2020multibody,hickok2022bounded}, and many others. Researchers have studied models of opinion dynamics on social networks to gain insight into phenomena such as the propagation of false or misleading information \cite{srivastava2016computing, toccaceli2020opinion}, the emergence of consensus opinions \cite{maletic2014consensus,neuhauser2020multibody, sahasrabuddhe2021modelling}, and the formation of echo chambers \cite{baumann2020modeling,brooks2023}. See \cite{sirbu2017opinion,noorazar2020classical} for reviews of opinion models.

Models of opinion dynamics necessarily involve many assumptions about the nature of the opinions of individuals, {the} interactions between individuals, and how such interactions affect the opinions of other individuals~\cite{olsson2023}. Most opinion models assume that agent opinions change {as a result of pairwise (i.e., dyadic) interactions.} {In opinion models, the agent opinions} are typically either real-valued scalars or real-valued vectors (e.g., if one wants to simultaneously model opinions {of} multiple things). In some models, the opinions have discrete values; in others, they have continuous values, such as in an interval of the real {line~\cite{noorazar2020classical}.} {Researchers typically consider discrete-value opinions when examining phenomena in which entities make discrete choices, such as when people vote for a candidate for a political office. By contrast, researchers often consider continuous-valued opinions when they want to explicitly account for a wide spectrum of views, such as political outlooks that range from very liberal to very conservative.} There are a wealth of opinion models \cite{xie2016review,noorazar2020classical}, which researchers study on networks to examine how social structures affect opinion dynamics. Examples of opinion models include the \nohyphens{DeGroot} consensus model \cite{degroot1974reaching}, voter models \cite{clifford1973model,holley1975ergodic} and their generalizations \cite{redner2019}, majority-rule models \cite{Galam02}, and bounded-confidence models \cite{bernardo2024}.

\begin{figure*}
    \includegraphics[width = \linewidth]{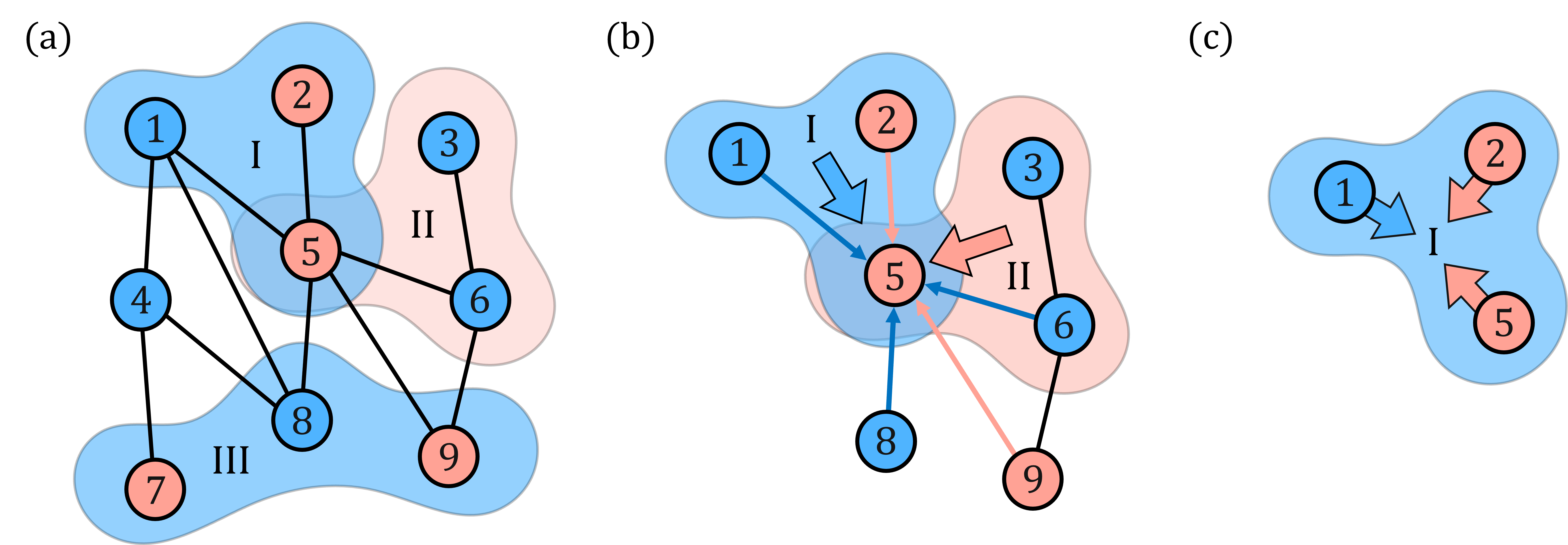}
    \caption{A schematic illustration of {how the opinions of nodes and groups are influenced by the opinions of other nodes and groups in our model}. (a) A hypergraph with 9 nodes and 3 groups. Nodes 1, 3, 4, 6, and 8 have opinion 1 {(in blue)}, and nodes 2, 5, 7, and 9 have opinion 0 {(in light red)}. Groups I and III have opinion 1 {(in blue)}, and group II has opinion 0 {(in light red)}. (b) Node 5's opinion is influenced by the opinion of its neighboring nodes 1, 2, 6, 8, and 9 {(thin arrows)} and by the opinions of groups I and II {(thick arrows)}. (c) Group I's opinion is influenced by the opinions of its constituent nodes 1, 2, and 5 {(thick arrows)}.}
    \label{model_example}
\end{figure*}

A basic assumption in most opinion models is that only the individual agents, which are represented by the nodes of a network, are endowed with an opinion. In the present paper, we relax this assumption by allowing groups of nodes to hold a collective opinion. Social groups, which range {in scope} from {family} and friendship units to large political and corporate organizations, help shape the fabric of society \cite{breiger1974duality,burningham1995individual,fine2014hinge}. In many situations, it is reasonable to posit that social groups themselves can possess opinions. For example, large corporate organizations sometimes take public stances on social issues \cite{eilert2020activist}. Additionally, courts such as the United States Supreme Court hear cases and document collective opinions through their decisions on these cases \cite{black2016us}. Moreover, a mathematics department at a university may broadcast a collective opinion, such as through documentation on its website or {through a hiring} decision, that differs markedly from the opinions of its individual faculty. The opinion of a research group, such as the applied-mathematics group, may also differ from the individual opinions of members of that group. In all of these examples, a group's opinion does not necessarily reflect a consensus among the members of that group, as {individuals in} the group can disagree (sometimes rather strongly) with the group opinion. Importantly, {groups} and individual members of a group can influence each other. Group opinions are influenced by the opinions of its constituent members, and the opinions of individuals are influenced by the {collective} opinions of the groups {in} which they participate \cite{cialdini2004social,griskevicius2006going}. 

In the present paper, we formulate and analyze a model of opinion dynamics in which both a network's nodes and its groups of nodes have {binary opinions.} The opinions of individual nodes are affected both by their neighbors in a network and by the opinions of the groups {in} which they participate. The opinions of groups are affected by the opinions of {their} constituent nodes. For simplicity, we neglect interactions between distinct groups. In Fig.~\ref{model_example}, we give a schematic illustration of how the opinions of nodes and groups are influenced by the opinions of other nodes and groups in our model, which we describe in detail in Sec.~\ref{our-model}. 
In our model, group opinions {can lead to oscillatory dynamics or excitable dynamics.} In the oscillatory regime, the mean opinion of a network develops self-sustained oscillations. In the excitable regime, {it is possible for} a situation in which most of the nodes have the same opinion {to} quickly and temporarily change to a situation in which most {of the} nodes have the other opinion. The emergence of {oscillatory and excitable} regimes depends strongly on the correlation between group structure and dyadic network adjacencies.
We develop a mean-field approximation of our model's dynamics and use it to gain insight into its dynamics. We find that the onset of excitable dynamics occurs due to a bifurcation from a stable equilibrium {to} sustained oscillatory dynamics. This bifurcation resembles the onset of excitable dynamics in models of neuronal dynamics \cite{sepulchre2018excitable,izhikevich2007dynamical}. {There is some qualitative similarity between the excitable dynamics in our model and the} formation of social fads, which is a collective behavior in which a topic, object, or behavior experiences an increase in popularity that forms suddenly, last{s} a short amount of time, and declines rapidly \cite{best2006flavor, aguirre1988collective}.

The role of group interactions, which are often called ``higher-order interactions" {or} ``polyadic interactions'', in opinion dynamics and other dynamical processes on networks has received much attention in the past few years~{\cite{battiston2020networks,battiston2021,bick2023,gao2023dynamics}}. The renewed interest {in} polyadic interactions in complex systems has built on foundational work that dates back many decades \cite{atkin1977combinatorial, abrams1983arguments, wilbur1990experimental, schaffer1984adult}. In social systems, {a lot of recent research has extended existing} models to incorporate polyadic interactions \cite{maletic2014consensus,iacopini2019simplicial, neuhauser2020multibody, neuhauser2021opinion, horstmeyer2020adaptive, cencetti2021temporal, sahasrabuddhe2021modelling, hickok2022bounded,noonan2021dynamics, goh2024, hebert2022source, st2024nonlinear,st2024paradoxes}. In these extensions, the opinions are associated with individual agents and they change due to both pairwise and  group interactions. By contrast, in our {model,} we also assign opinions to the groups themselves. In this sense, our work is {related} to recent studies of the synchronization of quantities that are defined on the edges and higher-order simplices of a simplicial complex {\cite{millan2025}.}

The incorporation of polyadic interactions into network dynamics can significantly affect qualitative behavior. For example, ``opinion jumping'' can occur in polyadic bounded-confidence models \cite{hickok2022bounded}{,} bistable regions can arise in polyadic models of disease spread \cite{iacopini2019simplicial, cisneros2021multi, landry2020effect}{, and the qualitative nature of synchronization transitions of phase oscillators are affected by} polyadic interactions \cite{skardal2020higher,adhikari2023synchronization}. Our observation that group opinions can induce oscillatory and excitable dynamics provides another example of how polyadic interactions can fundamentally modify the dynamics of networked systems. {For simplicity, we consider polyadic interactions only through groups of size 3 (i.e., the groups always have exactly 3 nodes). Although this choice prevents us from exploring the effects of group-size heterogeneity (which occurs in most social systems), it has the important advantage of allowing us {to} easily engage with our primary focus, which is to study the effects of {group} opinions. Even with this major assumption, our model has very rich dynamics that differ markedly from existing opinion models (which do not incorporate group opinions).}

{In our model, the} individual nodes do not account for their own current opinions when they update their opinions. With this choice of opinion updating, we emphasize the role of (both dyadic and polyadic) social interactions in opinion evolution. Essentially, we are examining a regime in which the effects of self-influence are small in comparison to the effects of social {interactions}. In this respect, our opinion model is very different from many opinion models \cite{noorazar2020classical}, but it follows the tradition of classical voter models \cite{clifford1973model,holley1975ergodic,redner2019}. {In our {model}, groups can influence themselves, but distinct groups cannot influence each other.} {Because of group self-influence, we can use our model to explore} the effects of social phenomena, such as ``pluralistic ignorance" \cite{prentice1996pluralistic, miller1987pluralistic} and ``groupthink" \cite{janis2008groupthink}, that can {decelerate} changes in group opinions. In pluralistic ignorance, the members of a group {believe incorrectly} that they hold a minority opinion within a group \cite{prentice1996pluralistic, miller1987pluralistic}. Pluralistic ignorance, which one can view as a ``minority illusion" \cite{kureh2020} in social dynamics, provides a potential mechanism to slow social change through a process of self-silencing, such that individuals appear to conform to a belief that they believe is held by the rest of a group \cite{bicchieri2014norms}. Groupthink refers to the {tendency} of individuals in a group to seek social conformity and thereby disregard their own opinions. Our opinion model does not directly encode pluralistic ignorance or groupthink, but one can view its group self-influence term (which {slows} changes in group opinions) as incorporating them indirectly.

Our paper proceeds as follows. {In Sec.~\ref{our-model}, we introduce our stochastic model of opinion dynamics. In Sec.~\ref{twoB}, we discuss the model of random hypergraphs on which we simulate these opinion dynamics.} In Sec.~\ref{three}, {we derive a mean-field description of our stochastic opinion model, study its steady-state behavior, and compare that behavior to the steady-state behavior of the original opinion model. We also discuss the assumptions and approximations that we use to derive the mean-field description.} In Sec.~\ref{four}, we discuss the formation of group--node discordance states. In Sec~\ref{five}, we examine the formation of excitable and oscillatory dynamics. Finally, in Sec.~\ref{six}, we summarize and discuss our findings. Our code, figures, {and data} are available at \url{https://github.com/CorbitSampson/Oscillatory_Excitable_Opinion_Dynamics.git}. 


\section{Our stochastic opinion model}\label{our-model}

In this section, we describe our {stochastic} model of opinion dynamics with group opinions. We consider a set {$\mathcal{V}$} of $N$ nodes, which we index by $i \in \{1,2,\ldots, N\}$. Each node holds a binary opinion, which is either $0$ or $1$. Our model uses discrete time, so the time $t \in \{0,1,\ldots\}$. Let $x_i^{\, t}$ denote the opinion of node $i$ at time $t$. As in {opinion models} on ordinary graphs, which assume dyadic interactions between nodes, our nodes are adjacent to each other if there is an edge between them in a graph {$\mathcal{G}$}, which has an associated adjacency matrix $A$. We assume that {$\mathcal{G}$} is undirected and unweighted, so $A_{ij} = A_{ji} =1$ if nodes $i$ and $j$ are adjacent {to each other} and $A_{ij} = A_{ji} = 0$ if they are not adjacent. We also {suppose that there are $S$ groups of nodes; each group is a subset of $\mathcal{V}$.} For simplicity, we suppose that all groups have exactly 3 nodes{. Each group has either opinion $0$ or opinion $1$.} We label the groups with the index $j \in \{1,2,\ldots, S\}$, and we denote the opinion of group $j$ at time $t$ by $y_j^{\, t}$. The $N \times S$ incidence matrix $M$ has entries $M_{ij} = 1$ if node $i$ participates in group $j$ and $M_{ij} = 0$ if it does not. We henceforth refer to {the} groups in our networks as ``triangles". {The dyadic degree of each node $i$ is its number of edges, and its triadic degree is its number of triangles. Node $i$'s dyadic degree is thus $k_i = \sum_j A_{ij}$, and its triadic degree is $q_i = \sum_j M_{ij}$. The mean number of edges per node is $\langle k \rangle = \sum_{i}k_{i}/N$, and the mean number of triangles per node is $\langle q \rangle = \sum_{i}q_{i}/N$.}
 
{The} opinion of a node is influenced both by the opinions of {its} adjacent nodes (i.e., dyadic influence) and by the opinions of the groups in which it participates (i.e., polyadic influence) [see Fig.~\ref{model_example}(b)]. The opinion of a group is influenced by the opinions of its constituent nodes [see Fig.~\ref{model_example}(c)]. {Groups do not directly influence other groups.}

 We now describe our model in detail. The opinions of the nodes {(i.e., individuals)} and triangles {(i.e., groups)} evolve stochastically according to the update rule
\begin{align}
    x_i^{\, t + 1} &= \begin{cases}
        1 & \text{with probability\,} p_i^N\left( {\bf x}^{\, t}, {\bf y}^{\, t}\right)  \\
        0 & \text{otherwise\,,}
    \end{cases}\label{node_model} \\
    y_j^{\, t + 1} &= \begin{cases}
        1 & \text{with probability\,} p_j^E\left({\bf x}^{\, t}, {\bf y}^{\, t}\right)  \\
        0 & \text{otherwise\,,}
    \end{cases}\label{edge_model}
\end{align}
where ${\bf x}^{\, t}=[x_1^t,x_2^t,\ldots, x_N^t]^{{\text T}}$ is a node opinion vector at time $t$ and ${\bf y}^{\,t} = [y_1^t,y_2^t,\ldots, y_{S}^t]^{\text T}$ is a triangle opinion vector at time $t$. 

Suppose that the probability that node $i$ adopts opinion 1 is a {nonlinear} function of a linear combination of the opinions of its adjacent nodes (via its incident edges) and the opinions of the triangles {in} which it participates. That is,
\begin{equation} \label{node_prob}
    p_i^N\left({\bf x}^{\, t},{\bf y}^{\,t}\right) = f_N\left(a \bar{x}_i^{\, t} + b \bar{y}_i^{\, t} \right)\, ,
\end{equation}
where
\begin{align}
    \bar{x}_i^t &= \sum _{j = 1}^NA_{ij}x_j^t/\langle k\rangle \, ,\label{xbar} \\
    \bar{y}_j^t &= \sum _{k = 1}^{S} M_{jk}y_k^t/\langle q\rangle \, ,\label{ybar} 
\end{align}
the {influence} function $f_N$ is a {sigmoidal} function {(see Eq.~\eqref{sigf} below)}, and $a$ and $b$ are real-valued constants \footnote{One can absorb the quantities $\langle k \rangle$ and $\langle q \rangle$ in Eqs.~\eqref{xbar} and \eqref{ybar} into the model parameters $a$, $b$, $c$, and $d$.}. The parameter $a$ encodes the influence of node opinions on their neighbors, and the parameter $b$ encodes the influence of triangle opinions on their constituent nodes, as we will discuss in more detail below. As {we} discussed in Sec.~\ref{sec:level1}, nodes do not consider their own current opinions when they update their opinions (i.e., {$A_{ii} = 0$} for all $i$){.}

{We suppose that} the probability that a triangle adopts opinion 1 is a {sigmoidal} function of a linear combination {of its own current opinion and the opinions of its constituent nodes}. That is,
\begin{equation} \label{edge_prob}
    p_j^E\left({\bf x}^{\, t},{\bf y}^{\,t}\right) = f_E\left(c\bar{z}_j^{\, t} + d y_j^{\, t}\right)\, ,
\end{equation}
where
\begin{equation}
    \bar{z}_j^t = \frac{1}{3}\sum_{i = 1}^N M_{ij}x_i^t\, ,\label{zbar}
\end{equation}
{the influence function} $f_E$ is a {sigmoidal} function {(see Eq.~\eqref{sigf} below)}, and $c$ and $d$ are real-valued constants. The parameter $c$ encodes the influence of node opinions on the triangles to which they belong, and the parameter $d$ encodes the tendency of a triangle to maintain its opinion. For simplicity, we use the same sigmoidal function for the {influence} function for all nodes and all triangles{. We thus write}
\begin{align}
	{f(z) := f_N(z) = f_E(z) 
	= \frac{1}{2}\left[1 + \tanh(m(z - \mu))\right]\, ,} \label{sigf}
\end{align}
where $\mu$ is the inflection point of the sigmoid $f(z)$ and $1/m$ is proportional to the width of the sigmoid's transition region. We use a sigmoidal function because it is convenient for representing saturating interactions \cite{sigmoid2015}.
Researchers have used sigmoidal functions {in models of many other scenarios, including} echo chambers and polarization \cite{baumann2020modeling}, smooth bounded-confidence {dynamics} \cite{brooks2023}, and {many other} saturating interactions, which occur in diverse fields that range from neuroscience to robotics \cite{bizyaeva2022nonlinear}. 

{In Table~\ref{tab:table1}, we summarize the parameters and other key quantities of our model. These parameters include the influence parameters ($a$, $b$, $c$, and $d$), the influence function $f$ and its parameters $\mu$ and $1/m$, and parameters and other descriptors of network structure.}

{{It is important to highlight the parameters $a$, $b$, $c$, and $d$, which} encode the amount of (positive or negative) opinion influence at each time step. We interpret positive values of $a$, $b$, $c$, and $d$ as conforming to influence, and we interpret negative values of these parameters as rejecting influence.} The parameters $a$ and $b$ encode how much nodes are influenced by their neighboring nodes (the parameter $a$) and by the groups in which they participate (the parameter $b$). In particular, $a > 0$ (respectively, $a < 0$) increases (respectively, decreases) the probability that a node transitions to or maintains opinion 1 when more of its neighbors have opinion 1. Analogously, $b > 0$ (respectively, $b < 0$) increases (respectively, decreases) the probability that a node transitions to or maintains opinion 1 as it participates in more groups with opinion 1. {The parameter $c$ plays an analogous role for group opinions as $b$ does for {node opinions. Specifically,} $c > 0$ (respectively, $c < 0$) increases (respectively, decreases) the probability that a group transitions to or maintains opinion 1 {when more of its participants have opinion 1.} The parameter $d$ encodes how {much} a group's current opinion affects its {subsequent opinion. In particular, $d > 0$ (respectively, $d < 0$) increases (respectively, decreases) the probability that a group maintains opinion $1$.} }

\begin{table}[t]
\caption{\label{tab:table1}
The parameters {and other key quantities} of our opinion model.
}
\begin{ruledtabular}
\begin{tabular}{l p{0.7\linewidth}}
\multicolumn{1}{c}{\textrm{Parameter}}&
\textrm{{Description}}\\
\colrule
$a$ & {Influence} of individuals on other individuals\\ \hline
$b$ & {Influence} of groups on individuals \\ \hline
$c$ & {Influence} of individuals on groups \\ \hline
$d$ & {Influence} of a {group} on itself \\ \hline
$f$ & Sigmoidal influence function \\ \hline
$\mu$ & Inflection point of the sigmoid $f$ \\ \hline
$1/m$ & Width of the transition of the sigmoid $f$ \\ \hline
$r$ & {Correlation} coefficient between dyadic degree and triadic degree \\ \hline
$\mathcal{P}$ & {Hyperdegree} distribution \\ \hline
$P_1$ & {Marginal} dyadic degree distribution \\ \hline
$P_2$ & {Marginal} triadic degree distribution \\ \hline
$k_i$ & {Dyadic degree (i.e., ordinary node degree) of node $i$}   \\ \hline
$q_i$ & Triadic degree of node $i$
\end{tabular}
\end{ruledtabular}
\end{table}


\section{Random-hypergraph model}\label{twoB}

It is convenient to use hypergraphs to describe our networks, which consist of nodes{, edges, and triangles.} A hypergraph is a generalization of a graph that includes both ordinary edges (i.e., dyadic adjacencies) and hyperedges with more than two nodes (i.e., polyadic adjacencies)~{\cite{newman2018networks,bick2023}}. Following standard convention, we refer to any of these adjacencies as hyperedges. Mathematically, a hypergraph {$\mathcal{H}_\mathcal{G} = (\mathcal{V},\mathcal{E})$} consists of a set {$\mathcal{V}$} of nodes and a set {$\mathcal{E}$} of hyperedges. Each hyperedge is a nonempty subset of {$\mathcal{V}$}; the number of nodes in this subset is the ``size'' of the hyperedge.

In an ordinary graph, each node {$i \in \mathcal{V}$} has an associated degree $k_{\, i}$, which indicates the number of edges that are attached (i.e., ``incident'') to it. In a hypergraph, the hyperdegree of node $i$ is the vector {${\bf k}_i = [k_{\, i}^{(2)},k_{\, i}^{(3)}, \ldots, k_{\, i}^{(L)}]$}, {where $L$} is the size of its largest hyperedge and the {$l$th-order} degree {$k_i^{(l)}$} is the number of {size-$l$} hyperedges that are incident to node {$i$}. Each hypergraph has a hyperdegree distribution $\mathcal{P}({\bf k})$, which encodes the probabilities that a uniformly-randomly-chosen node has hyperdegree ${\bf k}$ for each ${\bf k}$. {We consider hypergraphs with hyperedges of sizes $2$ and $3$; the hyperedges of size $2$ encode dyadic adjacencies, and the hyperedges of size $3$ encode the triadic (i.e., group) adjacencies. For such hypergraphs, the hyperdegree of a node is ${\bf k} = [k,q]$.}
 
To study our opinion model, it is convenient to use random hypergraphs with specified hyperdegree sequences{.} {{We use} such configuration-model random hypergraphs because we are able to control their hyperdegree sequences. In Appendix~\ref{randhypergraphs}, we provide a detailed discussion of the random-hypergraph model that we employ.} The formation of hyperedges in this random-hypergraph model depends only on the specified hyperdegree of each node, so one can employ hyperdegree-based compartmental models when studying dynamical processes on {the} hypergraphs that it generates. Such techniques have been used extensively in the study of disease spread on networks~\cite{kiss2017,ferraz2024}.

To {examine} the effects of {correlations} between the dyadic degree $k$ and {the} triadic degree $q$, we {employ} a convenient family of hyperdegree distributions {to produce the degree sequences in our random hypergraphs.} Given the marginal degree distributions $P_1(\cdot)$ and $P_2(\cdot)$ for the edges and triangles, respectively, there are two extremes for the joint distribution $\mathcal{P}(k,q)$. In one extreme, we {let $k = q$, which implies that $\mathcal{P}(k,q) = P_1(k)\delta(k - q)$}. In the other extreme, {$k$ and $q$ are uncorrelated, which implies that} $\mathcal{P}(k,q) = P_1(k)P_2(q)$. To systematically explore the effects of {correlations} between the dyadic and triadic degrees, we use a hyperdegree distribution that interpolates between these two extremes. This joint distribution is
\begin{equation} \label{interp_DD}
    \mathcal{P}(k,q) = P_1(k)P_2(q)(1 - r) + P_1(k)\delta(q - k)r\, ,
\end{equation}
where the Pearson correlation coefficient $r \in [0,1]$ between the dyadic and triadic degrees parameterizes the amount of correlation between these degrees. When $r = 0$, the dyadic degree $k$ and triadic degree $q$ are uncorrelated; when $r = 1$, we have that $k = q$ for every node.

To examine the effects of degree heterogeneity, {we} suppose that the degree distributions $P_1(k)$ and $P_2(q)$ have the approximate power-law form 
\begin{align}
    P_1(k) = P_2(k) = P(k) &= 
    \begin{cases}
    	    \left(\frac{\gamma - 1}{k_{ \text{min}}^{1 - \gamma}}\right)k^{-\gamma}\, , \quad k \geq k_{\text{min}} \\
    	    0\, , \qquad \qquad \quad \text{otherwise}\, ,
    \end{cases}\label{lim_DD}
\end{align}
where $k_{\text{min}}$ is the minimum degree. In a particular hypergraph that we construct using our random-hypergraph model, we generate the hyperdegree of each node using bivariate inverse sampling from the distribution that is described by Eqs.~(\ref{interp_DD}) and (\ref{lim_DD}). We then construct the hypergraph {using a procedure that we describe} in Appendix~\ref{randhypergraphs}.


{\section{Mean-field approximation, initial conditions, and steady-state solutions}}

{\subsection{Mean-field approximation of Eqs.~(\ref{node_model})--(\ref{edge_model})}
\label{three}}

We develop a mean-field description that approximates the dynamics of {our stochastic opinion model~(\ref{node_model})--(\ref{edge_model}).} Our mean-field description tracks the dynamics of three order parameters: 
\begin{enumerate}
\item{the expected fraction $V^t$ of nodes with opinion 1 in {a uniformly random selected edge} at time $t${;}}
\item{the expected fraction $U^t$ of nodes with opinion 1 in {a uniformly random selected triangle} at time $t${; and}} 
\item{the expected fraction $Y^t$ of triangles with opinion 1 at time $t${.}} 
\end{enumerate}
Alternatively, $V^t$ represents the probability {of moving} to an opinion-1 node by following an edge that one chooses uniformly at random.  

In this section, we present a simplified derivation of our mean-field approximation. {In Appendix~\ref{meanfield_app}, we show a detailed derivation of this approximation.} Because we generate hypergraphs using a configuration model, the probability that there is a hyperedge that connects a group of nodes depends only on the hyperdegrees of those nodes. Therefore, we can use a hyperdegree-based compartmental model to obtain mean-field equations. First, we approximate the order parameters $V^t$ and $U^t$ in terms of {the expected fraction $x_{\bf k}^t$ of} nodes with hyperdegree ${\bf k} = [k,q]$ that have opinion 1 at time $t$. {As we show} in Appendix~\ref{meanfield_app}, these approximations take the form
\begin{align}
    V^t &= \sum_{{\bf k}}\frac{k \mathcal{P}({\bf k})x^t_{{\bf k}}}{\langle k \rangle}\, ,\label{V}\\
  U^t &= \underset{{\bf k}}{\sum} \frac{q\mathcal{P} ({\bf k})x_{{\bf k}}^{t}}{\langle q  \rangle}\, .\label{U}
\end{align}
{The variables} $V^t$ and $U^t$ are {both} closely related to --- but can differ from --- the expected {fraction $\sum_{\bf k} \mathcal{P} ({\bf k})x_{{\bf k}}^{t}$ of nodes with opinion $1$ at time $t$.}

{To} obtain a system of discrete-time evolution equations for the order parameters {$V^t$, $U^t$, and $Y^t$,} consider the probability [see Eq.~\eqref{node_model}] that a node $i$ with hyperdegree ${\bf k} = [k,q]$ has opinion 1 at time $t + 1$. Assuming that all nodes with the same hyperdegree behave {in the same way, we seek to} approximate the variables $\bar{x}_i^{\, t}$, $\bar{y}_i^{\, t}$, $\bar{z}_j^{\, t}$, and $y_j^t$ that appear in the probabilities {(\ref{node_prob})--(\ref{zbar})} in terms of the order parameters. 

{We approximate $\bar{x}_i^{\, t}$, which is the normalized number of neighbors of node $i$ that have opinion 1, by} 
\begin{align}
    \bar{x}_i^{{\, t}} & \approx kV^t/\langle k\rangle\, . \label{xbarapprox}
\end{align}
The term $\bar{y}_i^{\, t}$, which is the normalized number of triangles that are attached to node $i$ {and} have opinion 1 [see Eq.~(\ref{zbar})], is approximately
\begin{align}\label{ybarapprox}
    \bar{y}_i^{{\, t}} \approx q Y^t/\langle q\rangle 
\end{align}
because node $i$ is attached to $q$ triangles and $Y^t$ is the expected fraction of triangles that have opinion 1. 

{We} insert the approximations (\ref{xbarapprox}) and (\ref{ybarapprox}) into Eqs.~(\ref{node_model})--(\ref{node_prob}) to obtain
\begin{align}
    x_{{\bf k}}^{t + 1} &\approx  f\left(\frac{ak}{\langle k \rangle}V^t + \frac{bq}{\langle q \rangle}Y^t \right)\, .\label{xkdyn}
\end{align}
{Under} the mean-field assumption that all triangles behave {in the same way} (i.e., $y_j^t = y^t$ and $\bar{z}_j^{{\, t}} = \bar{z}^{{\, t}}$ for all $j$), {the} time evolution of the expected fraction $Y^t$ of triangles with opinion 1 satisfies
\begin{align} \label{Ydymwithz}
    Y^{t + 1} = f(c \bar{z}^{\, t} + d)Y^t +f(c \bar{z}^{\, t})(1 - Y^t) \,.
\end{align}
{Similarly} to our approximation of {$\bar{y}_i^{\, t}$ in Eq.~(\ref{ybarapprox}), we approximate $\bar{z}^{\, t}$ by
\begin{align} 
    \bar{z}^{{\, t}}&\approx U^t \, , \label{zbarapprox}
\end{align}
which is the expected fraction of nodes at time $t$ with opinion 1 in a triangle that we select uniformly at random.}
Substituting Eq.~(\ref{zbarapprox}) into Eq.~(\ref{Ydymwithz}) yields
\begin{equation}
    Y^{\, t+1} = Y^tf\left(cU^t + d \right) + (1 - Y^t)f(cU^t)\, .\label{Yequ}
\end{equation}
{Inserting} Eq.~(\ref{xkdyn}) into Eqs.~(\ref{V})--(\ref{U}) yields a closed map for the time evolution of the three order parameters:
\begin{align} \label{meandyn}
\begin{split}
    V^{\, t + 1} &= \sum_{k}\sum_{q} \frac{k   \mathcal{P} (k,q)}{\langle k \rangle} f\left( \frac{a\,k}{\langle k \rangle}V^t + \frac{b\,q}{\langle q \rangle}Y^t \right)\, ,\\
    U^{\, t + 1} &= \sum_{k}\sum_{q} \frac{q  \mathcal{P} (k,q)}{\langle q \rangle} f\left( \frac{a\,k}{\langle k \rangle}V^t + \frac{b\,q}{\langle q \rangle}Y^t \right)\, ,\\
    Y^{\, t + 1} &= Y^tf\left(cU^t + d \right) + (1 - Y^t)f(cU^t)\, .
\end{split}
\end{align}

The mean-field description (\ref{meandyn}) relies on various approximations, which we now summarize and discuss. First, our mean-field description is a hyperdegree-based compartmental model, so it assumes that the expected time evolution of all nodes with hyperdegree ${\bf k}$ is the same. (For example, the probability that each such node has opinion 1 at time $t$ is $x_{\bf k}^t$.) This approximation relies on the fact that {we assume that all nodes of hyperdegree ${\bf k}$ possess} the same type and number of expected connections. Our mean-field description also assumes that {the dyadic and triadic degrees of each node are both sufficiently large that} we can replace the variables $\bar{x}_i^{{\, t}}$, $\bar{y}_i^{{\, t}}$, and $\bar{z}_i^{{\, t}}$ by their means [as we did in Eqs.~(\ref{xbarapprox}), (\ref{ybarapprox}), and (\ref{zbarapprox})]. In particular, we do not expect our mean-field approximation to give a good approximation for sparse hypergraphs. One can generalize our mean-field description to account for hypergraph models (e.g., degree-assortative random hypergraphs \cite{chodrow2020configuration}) in which nodes have intrinsic variables and connect to each other with probabilities that depend on these variables. Such generalizations of configuration models have a long history of success in investigations of dynamical processes on graphs \cite{melnik2014}.


{\subsection{Selection of initial conditions}
\label{intextIC}

{We now discuss our selection of initial conditions of our stochastic opinion model~\eqref{node_model}--\eqref{edge_model} and its mean-field approximation~\eqref{meandyn}. The spaces of initial conditions for these two models differ drastically from each other, so we need to select initial conditions that allow us to compare these models as effectively as possible.} 

In our simulations of the stochastic opinion model (\ref{node_model})--(\ref{edge_model}), the initial opinion of each node and each triangle is either $0$ or $1$. To reduce the number of parameters {that we need} to describe the initial conditions of the stochastic opinion model, we {specify} only the initial probabilities that nodes and triangles {that} have opinion $1$. {We specify the initial opinion of each node and each triangle independently. Each node initially has opinion $1$ with probability $u_1$ and opinion $0$ with probability $1 - u_1$, and each triangle initially has opinion $1$ with probability $u_2$ and opinion $0$ with probability $1 - u_2$. {We} then} specify the initial conditions of the stochastic opinion model (\ref{node_model})--(\ref{edge_model}) as ordered pairs $(u_1,u_2)$ {of initial probabilities}.

{The mean-field map~\eqref{meandyn} describes} the evolution of the three orders parameters $V^t$, $U^t$, and $Y^t$, {so we directly} specify the initial conditions {$V^0$, $U^0$, and $Y^0$} of these order parameters. {This specification contrasts with our selection of initial conditions in the stochastic opinion model (\ref{node_model})--(\ref{edge_model}), for which we can specify initial conditions using only two parameters.} {To mitigate this discrepancy}, in {our} examples that compare simulations of (\ref{node_model})--(\ref{edge_model}) and \eqref{meandyn}, we let $V^0 = U^0 = u_1$ and $Y^0 = u_2$ {to} make the initial conditions {for the two descriptions} as similar as possible. We make this choice because the order parameters $V^t$ and $U^t$ are related (but not equal) to the fraction of nodes with opinion 1 [see Eqs.~(\ref{V}) and (\ref{U})] and $Y^t$ is the expected fraction of triangles with opinion $1$.

{In several of our examples, we compare} many possible choices of the initial conditions of the stochastic opinion model (\ref{node_model})--(\ref{edge_model}) and mean-field map \eqref{meandyn}. {In Appendix~\ref{ICS}, we give a detailed description of how we select these initial conditions.}}


\subsection{Steady-state solutions of {the stochastic opinion model}~(\ref{node_model})--(\ref{edge_model})}

We now examine the steady-state solutions (i.e., states in which the {order parameters} $V^t$, $U^t$, and $Y^t$ are constant) of the stochastic opinion model~(\ref{node_model})--(\ref{edge_model}) by studying the fixed points of the mean-field equations (\ref{meandyn}). We obtain qualitatively similar steady-state solutions and bifurcations for any value of the hyperdegree correlation $r \in [0,1]$. Therefore, for simplicity, we assume in the present discussion that $r = 1$, which implies that the dyadic degree and triadic degree are equal (i.e., $k = q$). We use this assumption for the remainder of this section and throughout Sec.~\ref{four}. We will see in Sec.~\ref{five} that relaxing this assumption results in qualitatively different {dynamics.} Under this assumption, $V^t = U^t$ and the mean-field equations~(\ref{meandyn}) reduce to 
\begin{align} \label{reducedmeandyn}
\begin{split}
    V^{t + 1} &= \sum_k \frac{k P(k)}{\langle k \rangle} f\left( \frac{k}{\langle k \rangle} \left( a V^{t} + b Y^{t}\right) \right)\, ,\\
    Y^{t + 1} &= Y^{t}f\left(cV^{t} + d\right) + (1 - Y^{t})f\left(cV^{t}\right)\, .
\end{split}
\end{align}
Any fixed point $[V^t, Y^t] = [V^*,Y^*]$ of the map (\ref{reducedmeandyn}) must satisfy
\begin{align}
    V^{*} &= \sum_k \frac{k P(k)}{\langle k \rangle} f\left( \frac{k}{\langle k \rangle} \left(a V^{*} + b Y^{*}\right) \right)\, ,\label{reduceV*}\\
    Y^{*} &= Y^{*}f\left(cV^{*} + d\right) + (1-Y^{*})f\left(cV^{*}\right)\, . \label{reduceY*}
\end{align}
Solving (\ref{reduceY*}) for $Y^*$ and substituting the result into Eq.~(\ref{reduceV*}) shows that the fixed points of Eqs.~(\ref{reducedmeandyn}) have the form $[V^*,Y^*] = [F(V^*), G(V^*)]$, where
\begin{align} \label{FG}
\begin{split}
    F(V) &=\sum_k \frac{k P(k)}{\langle k \rangle} f\left( \frac{k}{\langle k \rangle} \left( a V+ bG(V) \right) \right)\, ,\\
    G(V) &=  \frac{f(cV)}{1 + f(cV) - f(cV + d)}\,.
\end{split}
\end{align}
The equation $V^* = F(V^*)$ is a one-dimensional equation for $V^*$ that {one can solve} using a root-finding algorithm. After determining $V^*$, we obtain $Y^*$ using the equation $Y^* = G(V^*)$. 

\begin{figure}[t]
        \includegraphics[width = \linewidth]{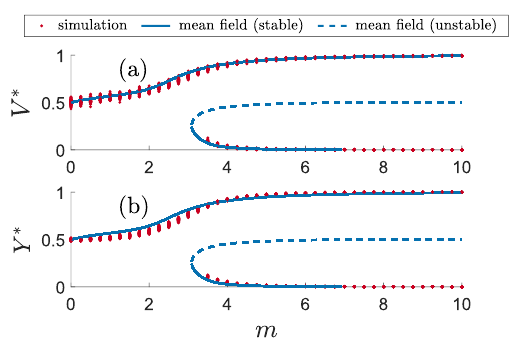}
        \caption{An example of a bifurcation of the {steady-state solutions} {when the dyadic and triadic degrees are {equal} (i.e., $r = 1$)} in simulations of our stochastic opinion {model~(\ref{node_model})--(\ref{edge_model})} and solutions of {the mean-field} equations~(\ref{reducedmeandyn}) for the parameter values {$a = b = c = d = \mu = 0.5$, {power-law exponent} $\gamma = 4$, and {mean degrees} $\langle k \rangle = \langle q \rangle = 20$}. {The bifurcation parameter is the inverse-width parameter $m$ of the sigmoidal influence function~\eqref{sigf}.} We show the values of (a) $V^*$ and (b) $Y^*$ that we obtain from the mean-field equations (solid and dashed curves) and from means of 100 simulations of our stochastic opinion model (dots).}
        \label{bifurcationM}
\end{figure}

To illustrate the usefulness of Eqs.~(\ref{reduceV*})--(\ref{reduceY*}) to study steady-state solutions of the {stochastic} opinion model~(\ref{node_model})--(\ref{edge_model}), we compare the fixed points that we obtain from the solution of Eqs.~(\ref{reduceV*})--(\ref{reduceY*}) with the results of simulations of Eqs.~(\ref{node_model})--(\ref{edge_model}). In Fig.~\ref{bifurcationM}, we plot both the steady-state values $V^*$ and $Y^*$ that we obtain from simulations of Eqs.~(\ref{node_model})--(\ref{edge_model}) (dots) and the fixed-point solutions of Eqs.~(\ref{reduceV*})--(\ref{reduceY*}) (solid and dashed curves) as a function of the sigmoid inverse-width parameter $m$ for $a = b = c = d = \mu = 1/2${,} power-law exponent $\gamma = 4${, and mean degrees $\langle k \rangle = \langle q \rangle = 20$}. Because $a$, $b$, $c$, and $d$ all have positive values, all nodes and groups experience only conforming influence. For each value of $m$, we iterate Eqs.~(\ref{node_model})--(\ref{edge_model}) for 400 steps and plot the values of $V^*$ and $Y^*$ after the final step. We do 100 independent simulations of this process with {evenly spaced initial conditions in the unit square (see Appendix~\ref{ICS}).} For each value of $m$, we use the same hypergraph with $N = 700$ nodes. Both our simulations of the stochastic opinion model~(\ref{node_model})--(\ref{edge_model}) and our analysis of the {mean-field} approximation (\ref{reducedmeandyn}) illustrate that the system transitions from a {regime with a single steady state to a regime with two distinct steady states} as we increase $m$. Although we observe some quantitative differences between our mean-field {description and direct simulations of the original stochastic opinion model}, the fixed points of the mean-field equations are reasonably successful at approximating the steady-state solutions of the {stochastic model.} The bifurcation in Fig.~\ref{bifurcationM} gives an interesting example of the behavior of our {opinion} model. A similar bifurcation was also observed in another opinion model with sigmoidal interactions~\cite{bizyaeva2022nonlinear}. Therefore, we do not focus on such bifurcations of steady-state solutions in situations with equal dyadic and triadic degrees ($r = 1$). Instead, we investigate novel features that arise due to the presence of group opinions. In particular, we observe (1) states in which the mean node and mean group opinions are different and (2) excitable and oscillatory opinion dynamics. The oscillator dynamics (see Sec.~\ref{five}) arise only when the dyadic degree and triadic degree are not fully correlated (i.e., when $r < 1$).

\section{Group--node discordance}\label{four}

An important feature of our opinion model is that it admits solutions in which the mean opinion of the nodes differs significantly from the mean opinion of the groups. We refer to these solutions as {\it group--node discordance states}. These states can model situations in which a social organization (or other social group) has a different official stance than the individuals who comprise that organization. In our model, we measure the discordance of a solution by calculating 
 \begin{align}
     D(V^*,Y^*) = |V^* - Y^*|\, {.} \label{capD}
 \end{align}
 {A} group--node discordance state occurs when $D(V^*,Y^*) > 0$. {The maximum possible discordance is $D(V^*,Y^*) = 1$.}

\begin{figure}[t]
    \includegraphics[width = \linewidth]{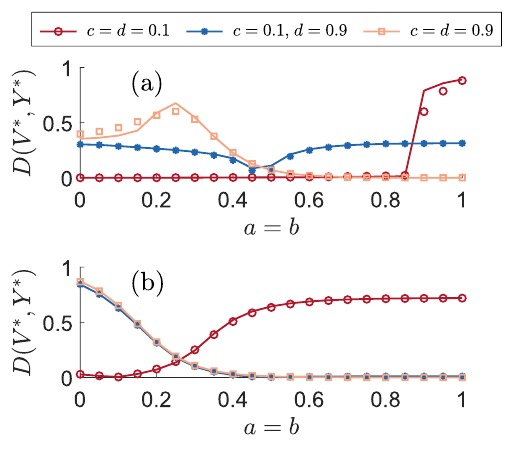}
    \caption{{The group--node discordance} $D(V^*,Y^*)$ versus the node-opinion influence parameter $a$, with $a = b$, for a single numerical solution $(V^*,Y^*) = (F(V^*),G(V^*))$ of Eqs.~(\ref{FG}) (solid curves) and the mean of $16$ independent simulations of the stochastic opinion model (\ref{node_model})--(\ref{edge_model}) for a single configuration-model hypergraph with $N = 2000$ nodes, {{inverse-width parameter} $m = 4$, {power-law exponent} $\gamma = 4$}, {mean degrees $\langle k \rangle = \langle q \rangle = 20$,} and several values of {the group-influence parameters} $c$ and $d$. We consider (a) $\mu = 0.5$ and (b) $\mu = 0.25$. {The initial conditions of the 16 simulations are evenly spaced in the unit square (see Appendix~\ref{ICS}).}} 
    \label{absweep}
\end{figure}

{We explore how
{group--node discordance states can arise for different strengths {of group influence}} (i.e., for different values of the group-influence parameters $c$ and $d$). We plot $D(V^*,Y^*)$ versus the node-opinion influence parameter $a$ (with $a = b$) {using} the mean {values} of $V^*$ and $Y^*$ {from} $16$ independent simulations of the stochastic opinion model (\ref{node_model})--(\ref{edge_model}), and {we compare this plot to} a numerical solution of the fixed-point equations (\ref{reduceV*})--(\ref{reduceY*}) for the mean-field approximation (\ref{reducedmeandyn}).} In {this} comparison, we use a single realization of a configuration-model hypergraph with $N = 2000$ nodes, {{inverse-width parameter} $m = 4$, {power-law exponent} $\gamma = 4$}, {and {mean degrees} $\langle k \rangle = \langle q \rangle = 20$} for both $\mu = 0.5$ [see Fig.~\ref{absweep}(a)] and $\mu = 0.25$ [see Fig.~\ref{absweep}(b)]. The initial {conditions of} the $16$ independent simulations are evenly spaced in the unit square ({see} Appendix~\ref{ICS}). For $\mu = 0.5$, we obtain more group--node discordance [i.e., larger values of $D(V^*,Y^*)$] when the node parameters $a$ and $b$ are very different from the hyperedge parameters $c$ and $d$. We see this in Fig.~\ref{absweep}(a) for $c = d = 0.1$ (red solid curve and open circles) and $c = d = 0.9$ (orange solid curve and open squares). In both Fig.~\ref{absweep}(a) and Fig.~\ref{absweep}(b), the maximum discordance occurs when $a + b$ is on the opposite side of $\mu$ as $c + d$. When $c = 0.1$ and $d = 0.9$, we observe a more uniform discordance in the system, with a small decrease near $a = b = 0.5$. When $\mu = 0.25$ [see Fig.~\ref{absweep}(b)] for $c = d = 0.1$ and $c = d = 0.9$, we {observe} the same general trend. The group--node discordance states arise most prominently when $c + d < \mu < a + b$ or $a + b < \mu < c + d$. For $\mu = 0.5$ [see Fig.~\ref{absweep}(a)], the transition to a group--node discordance state for $c = d = 0.1$ {looks like it may} be discontinuous; for $\mu = 0.25$ [see Fig.~\ref{absweep}(b)], the transition to {a group--node discordance state} is continuous.

\begin{figure}[h]
    \includegraphics[width = \linewidth]{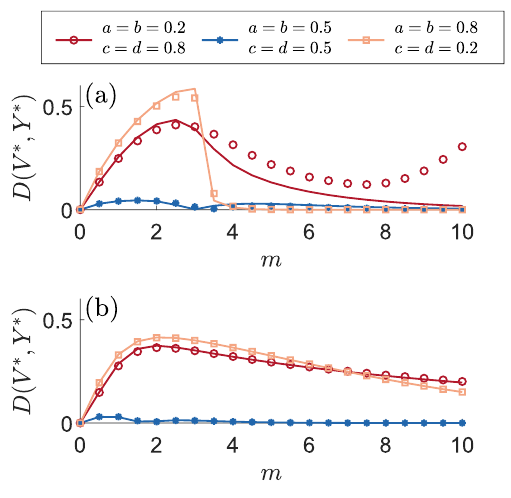}
    \caption{{The group--node discordance} $D(V^*,Y^*)$ versus the sigmoid {inverse-width} parameter $m$ for a single numerical solution $(V^*,Y^*) = (F(V^*),G(V^*))$ of Eqs.~(\ref{FG}) (solid curves) and the mean of $16$ independent simulations of the stochastic opinion model (\ref{node_model})--(\ref{edge_model}) for a single realization of a configuration-model hypergraph with $N = 2000$ nodes, {power-law exponent} $\gamma = 4$, {mean {degrees} $\langle k \rangle = \langle q \rangle = 20$,} and several values of $a$, $b$, $c$, and $d$. We consider (a) $\mu = 0.5$ and (b) $\mu = 0.25$. {The initial conditions of the 16 simulations are evenly spaced in the unit square (see Appendix~\ref{ICS}).}}
    \label{msweep}
\end{figure}

We also explore how the width (which is proportional to $1/m$) of the sigmoid transition region {affects} the onset of group--node discordance states {by calculating} $D(V^*,Y^*)$ versus $m$ for several values of $a$, $b$, $c$, and $d${. We show the results of our numerical simulations in Fig.~\ref{msweep}, which uses the same initial conditions, network parameters, and other conventions as Fig.~\ref{absweep}.} {When} $a = b = c = d = 0.5$ (blue curve and closed circles), the {width} parameter $m$ has a minimal effect and $D(V^*,Y^*)$ remains close to $0$, indicating {that there is} very little group--node discordance. For both $a = b = 0.2$, $c = d = 0.8$ (red curve and open circles) and $a = b = 0.8$, $c = d = 0.2$ (orange curve and open squares), the group--node discordance {$D(V^*,Y^*)$} has a maximum at an intermediate value of $m$. {We also observe} an interesting difference between the cases $\mu = 0.5$ [see Fig.~\ref{msweep}(a)] and $\mu = 0.25$ [see Fig.~\ref{msweep}(b)]. When $a = b = 0.8$ and $c = d = 0.2$, there is a {possibly} discontinuous transition from discordance to non-discordance for $\mu = 0.5$; however, the transition is continuous for $\mu = 0.25$.

\section{Excitable and oscillatory dynamics}\label{five}

Our opinion model also has excitable and oscillatory opinion dynamics. 
To illustrate these dynamics, we simulate both the stochastic opinion model (\ref{node_model})--(\ref{edge_model}) and the mean-field approximation (\ref{meandyn}) with the parameters $(a,b,c,d) = (1,-0.5,0.25,0.25)$ and $(m,\mu) = (8,0.25)$. 
In this regime, the nodes are influenced considerably by the opinions of their neighboring nodes ($a = 1$), nodes reject the opinions of their groups ($b = -0.5$), groups are influenced equally by their constituent nodes and their own opinions ($c = d = 0.25$), {and the sigmoidal influence function of the nodes and groups has a small inflection point ($\mu = 0.25$) and a very {steep} transition ($m = 8$).}

\begin{figure}[t]
\includegraphics[width = \linewidth]{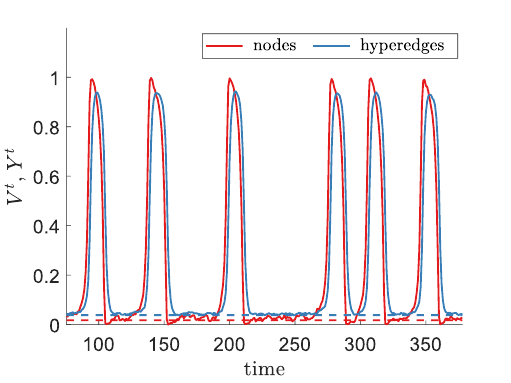}
    \caption{An example of {opinion} pulses in a single simulation of our stochastic opinion model (\ref{node_model})--(\ref{edge_model}) with parameter values $a = 1$, $b = -0.5$, $c = d = 0.25$, $\mu = 0.25$, and $m = 8$ for a configuration-model hypergraph with {equal} dyadic and triadic degrees (i.e., $r = 1$) that we draw from an approximate power-law distribution with exponent $\gamma = 4$ {and mean-degrees $\langle k \rangle = \langle q \rangle = 20$.} {We plot the expected node fraction $V^t$ in red and the expected triangle fraction $Y^t$ in blue.} The dashed lines show the fixed points that we obtain by solving Eqs.~(\ref{reduceV*})--(\ref{reduceY*}).}
    \label{simexamplepulse} 
\end{figure}

We first suppose that the dyadic and triadic degrees are equal ({so the correlation between them is $r = 1$}). This situation yields excitable dynamics, in which a dynamical system is initially at a locally stable steady-state solution, but --- for a sufficiently large perturbation (which is often called a ``stimulus") --- {it} experiences a large excursion through phase space before returning to the {steady-state solution} \cite{sepulchre2018excitable,izhikevich2007dynamical}. Excitable dynamics are common in neuronal and cardiac systems, and they are often associated with a system being near {a} bifurcation from a resting state to {sustained} spiking or oscillatory behavior \cite{sepulchre2018excitable,izhikevich2007dynamical,barrio2020excitable}. In Fig.~\ref{simexamplepulse}, we show the {order} parameters $V^t$ (red) and $Y^t$ (blue) from numerical simulations of Eqs.~(\ref{node_model})--(\ref{edge_model}) for the aforementioned parameter values and a configuration-model hypergraph with $N = 1000$ nodes and an approximate power-law hyperdegree distribution with $\gamma = 4$. The dashed curves show the fixed-point solution that we obtain by solving Eqs.~(\ref{reduceV*})--(\ref{reduceY*}). {In these simulations,} the expected node fraction $V^t$ remains close to the fixed-point solution for a short time before it increases sharply and then subsequently decreases and returns approximately to the fixed-point solution. The expected triangle fraction $Y^t$ has the same behavior; its dynamics follow $V^t$ with a short delay. We use the term {\it opinion pulses} for these spikes in $V^t$ and $Y^t$.

\begin{figure}[t]
\includegraphics[width = \linewidth]{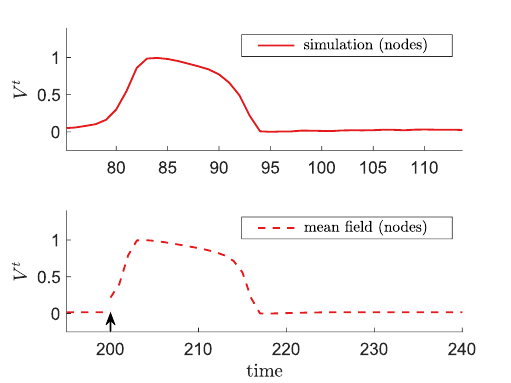}
    \caption{{Comparison of} a single pulse of {the expected node fraction} $V^t$ for (top) a single simulation of our stochastic opinion model (\ref{node_model})--(\ref{edge_model}) and (bottom) the mean-field equations (\ref{reducedmeandyn}) for the parameter values $a = 1$, $b = -0.5$, $c = d = 0.25$, $\mu = 0.25$, and $m = 8$ for a hypergraph with {equal} dyadic and triadic degrees {(i.e., $r=1$)} that we draw from an approximate power-law distribution with exponent $\gamma = 4$ {and mean-degrees $\langle k \rangle = \langle q \rangle = 20$.} We apply a stimulus (which is indicated by the black arrow) of $(\delta V, \delta U, \delta Y) = (0.2, 0.2, 0)$ to the mean-field equations to induce an excitation at time $t = 200$.}
    \label{examplepulse} 
\end{figure}

In Fig.~\ref{examplepulse}, we show $V^t$ as a function of time in simulations of {the stochastic opinion model}~(\ref{node_model})--(\ref{edge_model}) (top) and the mean-field equations (\ref{meandyn}) (bottom). For the mean-field equations, we introduce a stimulus at time 
$t = 200$ by increasing both {$V^t$} and {$U^t$} by 0.2 (vertical arrow). {The mean-field equations yield a single opinion pulse, which resembles the ones} that we observe in simulations of our stochastic opinion model (\ref{node_model})--(\ref{edge_model}), in which finite-size fluctuations seemingly provide a stimulus.

\begin{figure}[t]
\includegraphics[width = \linewidth]{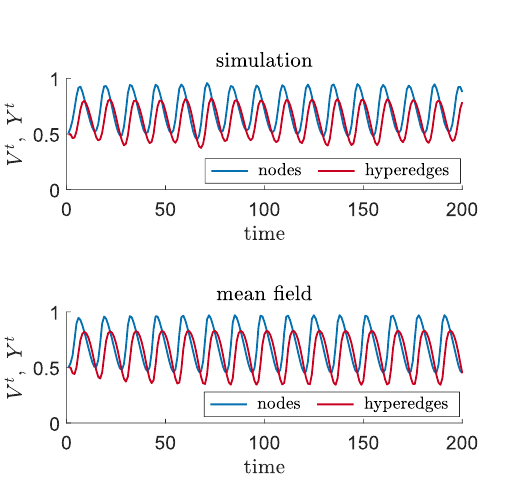}
    \caption{An example of the oscillatory dynamics in (top) a single simulation of our stochastic opinion model (\ref{node_model})--(\ref{edge_model}) and (bottom) the mean-field equations (\ref{meandyn}) for parameter values $a = 1$, $b = -0.5$, $c = d = 0.25$, $\mu = 0.25$, {$r = 0.15$, $m = 8$,} {$\gamma = 3.8$, and $\langle k \rangle = \langle q \rangle = 20$.} {We plot the expected node fraction $V^t$ in {blue} and the expected triangle fraction $Y^t$ in {red}.}
    The initial conditions of the stochastic opinion model are $(u_1,u_2) = (0.5,0.5)$, and the initial conditions of the mean-field equations are $(V^0,U^0,Y^0) = (0.5,0.5,0.5)$.}
    \label{oscexp}
\end{figure}

As we decrease the correlation $r$ between the dyadic and triadic degrees, the opinion pulses become more frequent until they eventually become self-sustained oscillations. {In Fig.~\ref{oscexp}, we show an example of such oscillations for $r = 0.15$.} {We again observe that the  mean-field equations \eqref{meandyn} successfully reproduce the qualitative behavior of the stochastic opinion model~(\ref{node_model})--(\ref{edge_model}).} 

\begin{figure}[t]
    \includegraphics[width = \linewidth]{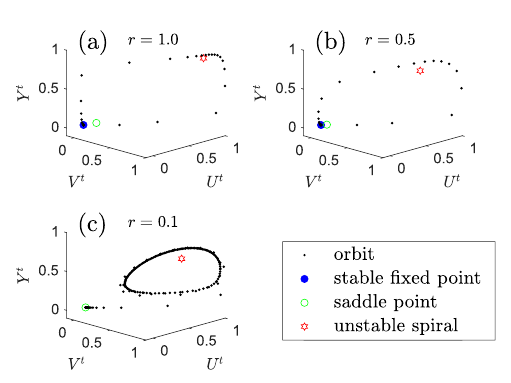}
    \caption{{The phase-space trajectories of solutions of the mean-field equations (\ref{meandyn}) with a perturbation $(\delta V,\delta U, \delta Y) = (0.25,0.25,0)$ for hyperdegree distributions with correlations} between the dyadic and triadic degrees of (a) $r = 1$, (b) $r = 0.5$, and (c) $r = 0.1$. The merging of stable and unstable fixed points near an unstable spiral leads to the transition from excitable to oscillatory dynamics. The {other} parameter values are $a = 1$, $b = -0.5$, $c = d = 0.25$, $\mu = 0.25$, $m = 8$, $\gamma = 4$, {and {$\langle k \rangle = \langle q\rangle = 20$.}}}
    \label{3d_map_pulse} 
\end{figure}

One can use the mean-field equations (\ref{meandyn}) to understand the transition from excitable to oscillatory dynamics as the correlation $r$ decreases. When $r = 1$, the mean-field map~(\ref{meandyn}) has three fixed points [see Fig.~\ref{3d_map_pulse}(a)]. We obtain these fixed points using root-finding methods and {determine their linear stability by calculating} the eigenvalues of the Jacobian matrix {of \eqref{meandyn}}. We compute the Jacobian matrix numerically using the ``Adaptive Robust Numerical Differentiation" package for MATLAB \cite{matlabpackage}. One of the fixed points [see the {closed blue circle} in Fig.~\ref{3d_map_pulse}(a,b)] is linearly stable. {It is located near the origin, so we {refer to it as} the ``near-0" fixed point.} A nearby fixed point [see the open green circle in Fig.~\ref{3d_map_pulse}(a)--(c)] is a saddle. The third fixed point [see the red star in Fig.~\ref{3d_map_pulse}(a)--(c)] is an unstable spiral{, and we refer to this spiral as the {``away-from-0"} fixed point.} When the mean-field system~(\ref{meandyn}) is {close to the near-0 fixed point (which is linearly stable)} and is perturbed so that it crosses the stable manifold of the saddle, it makes an excursion through phase space. It gets close to the unstable spiral and then returns to the stable fixed point, completing an opinion pulse. In Fig.~\ref{3d_map_pulse}(a), we show an example of such a trajectory in phase space. As we decrease $r$, the stable and saddle fixed points approach each other [see Fig.~\ref{3d_map_pulse}(b)], {and one can then create opinion pulses using smaller stimuli.} Eventually, {the linearly stable fixed point and the saddle} collide in a {SNIC ({i.e., saddle--node} on invariant circle) bifurcation,} resulting in oscillatory behavior [see Fig.~\ref{3d_map_pulse}(c)]. {See \cite{izhikevich2007dynamical} for details about SNIC bifurcations.} 

\begin{figure}[b]
\includegraphics[width = \linewidth]{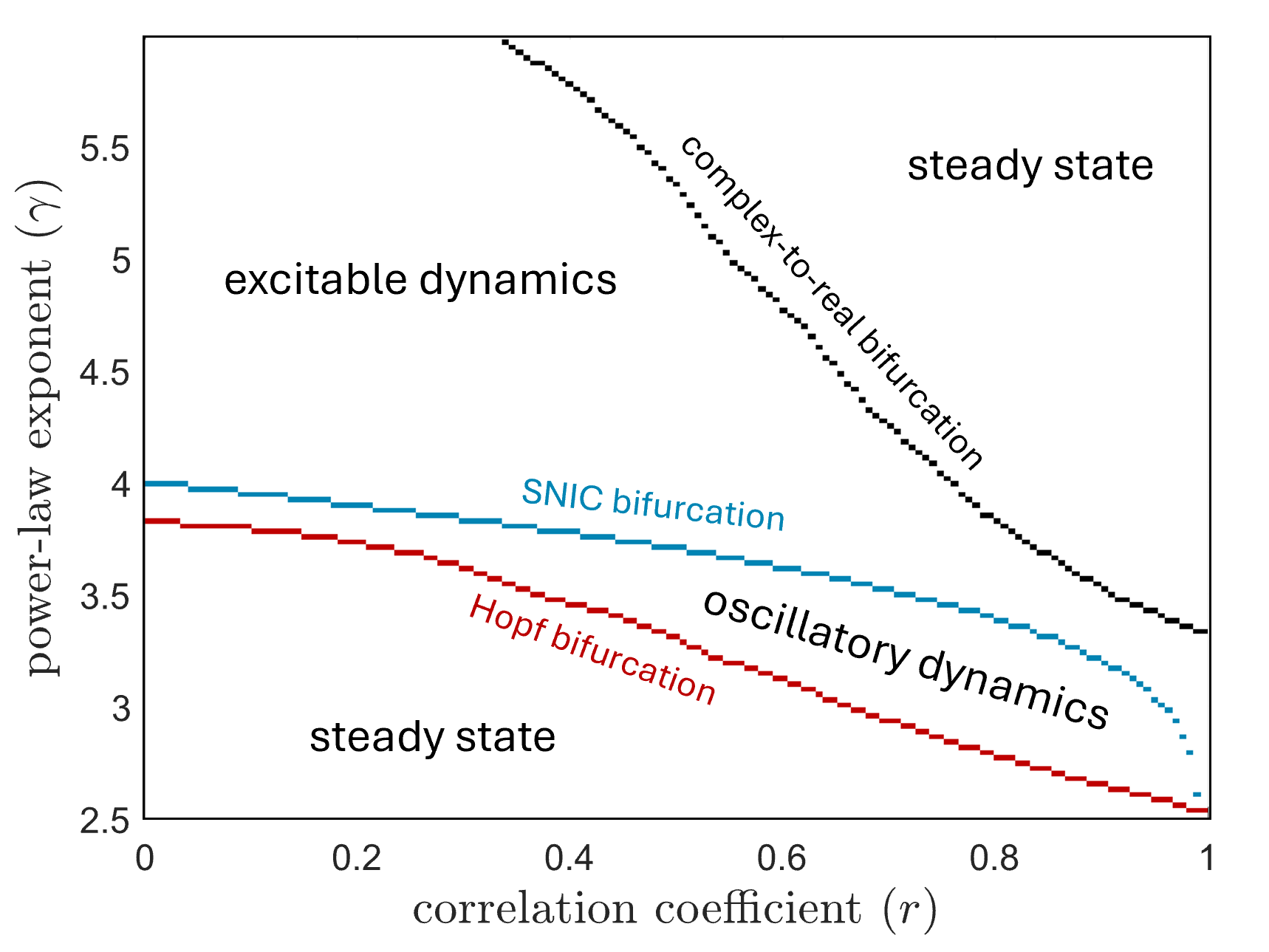}
    \caption{A bifurcation diagram of the transitions between stationary, excitable, and oscillatory states of the mean-field equations~\eqref{meandyn}. Each curve indicates a bifurcation in the {linear} stability of {the near-0 fixed point or the away-from-0 fixed point.} As one crosses the red curve, the {away-from-0} fixed point undergoes a Hopf bifurcation. As one crosses the light blue curve, the {near-0} fixed point undergoes a saddle--node bifurcation. As one crosses the black curve, the {away-from-0} fixed point transitions from a {linearly unstable spiral to a {linearly} unstable node.}}
    \label{wherebifurcation}
\end{figure}

{We have seen that oscillatory dynamics arise via a SNIC bifurcation as the correlation coefficient $r$ {decreases}. To obtain a broader perspective of the bifurcations and associated changes in qualitative dynamics that occur as the parameters change, we show a bifurcation diagram in $(r,\gamma)$ space in Fig.~\ref{wherebifurcation}. Recall that $\gamma$ is the exponent of the {approximate} power-law degree distributions of our configuration-model hypergraphs, so smaller values of $\gamma$ correspond to more heterogeneous degree distributions. For small values of $r$ and $\gamma$, the fixed points of the mean-field map (\ref{meandyn}) are linearly stable, which we indicate in the diagram by writing ``steady state''. Through the numerical linear stability analysis that we described above, we observe three bifurcations as $\gamma$ and $r$ increase (see Fig.~\ref{wherebifurcation}). We describe these three bifurcations in the order that they occur. First, there is a Hopf bifurcation of the {away-from-0} fixed point (red curve), which transitions from a {linearly} stable spiral to a {linearly} unstable spiral. This bifurcation marks the onset of oscillatory dynamics. Second, there is a SNIC bifurcation of the {near-0} fixed point (blue curve), which transitions from a saddle to a linearly stable fixed point.} {This bifurcation marks a change from oscillatory dynamics to excitable dynamics. Third, there is  a ``complex-to-real bifurcation" where the away-from-0 fixed point transitions from an unstable spiral to an unstable node. This bifurcation marks the return of steady-state behavior. The bifurcation diagram in Fig.~\ref{wherebifurcation} indicates where excitable and oscillatory dynamics occur in the parameter range $(r,\gamma)\in [0,1]\times[2.5,6]$.}

The mean-field {map~(\ref{meandyn}) is} an approximation of the stochastic opinion model (\ref{node_model})--(\ref{edge_model}), so we expect similar bifurcations to {occur} in the stochastic opinion model. Therefore, we believe that the {aforementioned bifurcations} {provide} a good explanation of the {onset of} excitable and oscillatory dynamics in both the mean-field {equations} (\ref{meandyn}) and the stochastic opinion model (\ref{node_model})--(\ref{edge_model}) {that they approximate}. To {numerically} verify this {conjecture}, we study the qualitative dynamics of the stochastic model {(\ref{node_model})--(\ref{edge_model})} for different values of both the correlation coefficient $r$ and the exponent $\gamma$ of the approximate power-law degree distribution. We compute the difference
\begin{equation} \label{scriptH}
    \mathcal{H}(V^t,\mathcal{I}) := \text{max}_{\mathcal{I}}(V^t) - \text{min}_{\mathcal{I}}(V^t)\, ,
\end{equation}
between the maximum and minimum values of {the expected node fraction} $V^t$ in an interval $\mathcal{I}$. A fixed-point solution of Eqs.~(\ref{meandyn}) gives $\mathcal{H}(V^t,\mathcal{I})\approx 0$ if we choose an interval $\mathcal{I}$ after the transient dynamics {disappear}. Oscillations and opinion pulses both yield $\mathcal{H}(V^t,\mathcal{I}) > 0$. In principle, one can also distinguish between pulses and oscillations by sliding {and/or varying the length of the interval $\mathcal{I}$,} but we {do not employ these approaches (and we have not examined them thoroughly).}

\begin{figure}[t]
\includegraphics[width = \linewidth]{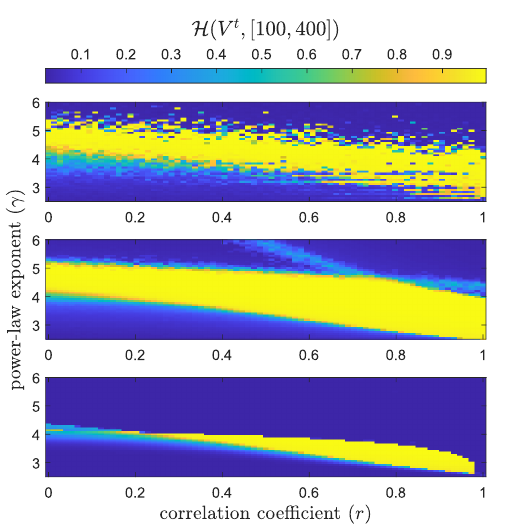}
    \caption{A heat map of $\mathcal{H}(V^t,\mathcal{I}) = \text{max}_{\mathcal{I}}(V^t) - \text{min}_{\mathcal{I}}(V^t)$ {[see Eq.~\eqref{scriptH}]} for (top) a mean of 25 independent simulations of the stochastic opinion model (\ref{node_model})--(\ref{edge_model}); (middle) {a} mean of $25$ independent simulations of the {stochastically perturbed mean-field equations~(\ref{dynsto}); and} (bottom) {a} mean of 25 independent simulations of the original mean-field equations (\ref{meandyn}). The horizontal axis is the correlation coefficient $r$, and the vertical axis is the power-law exponent $\gamma$. In these simulations, we use a single realization of a configuration-model hypergraph and the parameter values $a = 1$, $b = -0.5$, $c = d = 0.25$, $\mu = 0.25$, {$m = 8$, and $\langle k \rangle = \langle q \rangle = 20$.} For both the stochastic opinion model and the mean-field equations (both with and without stochastic fluctuations), we select $25$ evenly spaced initial conditions in the unit square (see Appendix~\ref{ICS}).}
    \label{wheredyn}
\end{figure}

In Fig.~\ref{wheredyn}, we plot $\mathcal{H}(V^t,\mathcal{I})$ versus the power-law exponent $\gamma$ from Eq.~\eqref{lim_DD} and the correlation coefficient $r$ for dyadic and triadic degrees in the interval $\mathcal{I} = [100,400]$, which seems to provide adequate time for the transient behavior to disappear. For each pair $(r,\gamma)$, we simulate the stochastic opinion model (\ref{node_model})--(\ref{edge_model}) and the mean-field equations (\ref{meandyn}) to obtain $\mathcal{H}$ from Eq.~(\ref{scriptH}). For both the stochastic opinion model and the mean-field equations{,} we select $25$ {evenly spaced initial conditions in the unit square} {(see Appendix~\ref{ICS})}. In the top panel {of Fig.~\ref{wheredyn}}, we plot $\mathcal{H}$ from simulations of the stochastic opinion model (\ref{node_model})--(\ref{edge_model}). In the bottom panel, we plot $\mathcal{H}$ from the mean-field equations (\ref{meandyn}). For a given value of the correlation coefficient $r$, {oscillatory or excitable} dynamics occur only for a narrow range of power-law exponents, illustrating that {these} dynamics are very sensitive to network structure. A stronger correlation between dyadic and triadic degrees (i.e., a larger $r$) requires a smaller value of $\gamma$ (i.e., a more heterogeneous network) for {oscillatory or excitable} dynamics to occur. Interestingly, perfectly correlated dyadic and triadic degrees (i.e., $r = 1$), which reduce the dimensionality of the mean-field equations (\ref{meandyn}) from $3$ to $2$, suppress the oscillatory dynamics. Observe that the yellow band [which indicates a large value of $\mathcal{H}(V^t,\mathcal{I})$] in the bottom panel of Fig.~\ref{wheredyn} does not extend to $r = 1$.

{Despite the qualitative similarities between the dynamics of the stochastic opinion model (\ref{node_model})--(\ref{edge_model}) {(see the top panel of Fig.~\ref{wheredyn})} and those of the mean-field equations \eqref{meandyn} {(see the bottom panel of Fig.~\ref{wheredyn})}, there are key differences between the qualitative dynamics of these models.} As we discussed above, although the mean-field equations {support} excitable dynamics, {they require a} stimulation to {yield} opinion pulses. To mimic the effect of {the stochastic model's finite-size fluctuations,} which are absent in the deterministic mean-field equations (\ref{meandyn}), we introduce a stochastic term. We consider the equations
\begin{align} \label{dynsto}
\begin{split}
    V^{\, t + 1} &= \sum_{k}\sum_{q} \frac{k   \mathcal{P} (k,q)}{\langle k \rangle} f\left( \frac{a\,k}{\langle k \rangle}V^t + \frac{b\,q}{\langle q \rangle}Y^t \right) + \sigma_1 ^t\, ,\\
    U^{\, t + 1} &= \sum_{k}\sum_{q} \frac{q  \mathcal{P} (k,q)}{\langle q \rangle} f\left( \frac{a\,k}{\langle k \rangle}V^t + \frac{b\,q}{\langle q \rangle}Y^t \right) + \sigma_2^t\, ,\\
    Y^{\, t + 1} &= Y^tf\left( cU^t + d \right) + (1 - Y^t)f(cU^t)\, .
\end{split}
\end{align}
In {Eqs.}~(\ref{dynsto}), we draw $\sigma_1^t$ and $\sigma_2^t$ uniformly at random from the interval $(0,Q)$ at each time step. These small stochastic perturbations act as a repeated small stimulus to the system. Our decision to not include a random stimulus to $Y^t$ arises from our observation that the fluctuations in the fraction of hyperedges with opinion 1 is smaller than the fluctuations in the fraction of nodes with opinion 1. This is the case because a hypergraph from the employed random-hypergraph model has many more hyperedges than nodes. {There} are $N \langle q \rangle/3$ hyperedges in an $N$-node hypergraph. 

In the middle panel of Fig.~\ref{wheredyn}, we show $\mathcal{H}$ {from} the {stochastically perturbed mean-field equations~(\ref{dynsto})} with an upper bound of {$Q \approx 0.036$} on the stochastic noise. We are not attempting to accurately reproduce the finite-size fluctuations of the stochastic opinion model (\ref{node_model})--(\ref{edge_model}). Instead, we seek to demonstrate that the mean-field description (\ref{meandyn}), when augmented with stochastic fluctuations, can produce excitable and oscillatory dynamics that are qualitatively similar to those that we obtain from simulations of the stochastic opinion model{.}

{By simultaneously examining Fig.~\ref{wherebifurcation} and Fig.~\ref{wheredyn},} we observe that the mean-field approximation {[see the bottom panel of Fig.~\ref{wheredyn}}] {has} a {region {with $\mathcal{H}(V^t,\mathcal{I}) > 0$} {of} a shape that is qualitatively similar {to}} the oscillatory region {in} Fig.~\ref{wherebifurcation}{.} {We expect this similarity because our simulations of the mean-field equations~\eqref{meandyn} in the bottom panel of Fig.~\ref{wheredyn} do not include perturbations or stimuli, and they thus do not exhibit excitable dynamics.} We also {observe} that the {region {with} $\mathcal{H}(V^t,\mathcal{I}) > 0$} {in} the {top and} middle panels of Fig.~\ref{wheredyn} is qualitatively similar to the {union of the} oscillatory {region and the lower part of the excitable region} {in} Fig.~\ref{wherebifurcation}. {When we use a larger value of $Q$ in the stochastically perturbed mean-field map \eqref{dynsto}, the region with $\mathcal{H}(V^t,\mathcal{I}) > 0$ more closely resembles the union of the entirety of the excitable and oscillatory regions in Fig.~\ref{wherebifurcation}.}


\section{Conclusions and discussion}\label{six}

We introduced and analyzed a stochastic model of opinion dynamics in which both nodes and groups of nodes have {binary opinions}. {This} opinion model includes novel dynamics that result directly from polyadic interactions. We showed that our model supports a richer repertoire of qualitative dynamics than {related} models in which only nodes have opinions. In particular, our opinion model has both excitable dynamics (in which brief but strong opinion swings arise from perturbations of a {steady-state solution}) and oscillatory dynamics (in which the mean opinions of the nodes and hyperedges have self-sustained oscillations). The excitable dynamics of our system {have} qualitative similarities with the dynamics of social fads \cite{best2006flavor}. In particular, {opinion changes appear} initially in a small number of individuals (or via {an} external perturbation for our mean-field approximation), experiences a surge that affects the majority of the system, and then quickly dies out. Our opinion model also possesses group--node discordance states, in which nodes and groups have contradictory opinions. {Our simulations of the stochastic opinion model and its} mean-field approximation both reveal that the excitable and oscillatory dynamics depend significantly on network structure ({specifically, on} dyadic degrees, polyadic degrees, and the correlation between them).

There are many interesting ways to extend our opinion model. As with all models of opinion dynamics, we greatly simplified human dynamics (or the dynamics between other animals) to formulate a mathematically and computationally tractable model that one can study systematically. For example, we assumed that opinions are binary {(instead of allowing more opinion states or continuous-valued opinions)}, that interactions occur through a known and time-independent hypergraph, and that {opinions evolve through precise mathematical rules.} It is worth relaxing {these} assumptions and exploring the consequences of doing so.

One {important} way to generalize our model is to incorporate various heterogeneities, including in the {group sizes, the interaction strengths (e.g., some groups or nodes may be more influential than others), and} the shapes of the sigmoidal functions (e.g., some nodes may be more likely than others to change their opinions). {For simplicity, we} limited our study to groups of size 3. {As we illustrated at length,} the dynamics that result from considering only groups of size 3 is already very rich. However, it is natural to expect that some phenomena occur only in networks with heterogeneous group sizes. For example, perhaps an opinion can propagate from small groups to large groups (or vice versa). Just as the degree distribution of a graph can significantly influence the qualitative behavior of dynamical processes on it \cite{newman2018networks,porter2016}, we expect that the hyperedge-size distribution (along with hyperdegree distributions) influences the qualitative behavior of dynamical processes on a hypergraph. {In our study,} we also neglected interactions between {distinct} groups, which are likely to introduce additional interesting dynamics. Another potentially interesting extension of our model is the inclusion of node self-influence, as individuals typically have some conviction in their prior beliefs. Additionally, although our mean-field approximation adequately reproduced the observed dynamics and provided some theoretical insights, it is based on the assumption that the hypergraph that describes the {nodes and the groups} is generated by a configuration model. It is worthwhile to extend our mean-field approximations to stochastic-block-model hypergraphs with assortative mixing (which can encode homophily) \cite{landry2022hypergraph} and community {structure \cite{landry2023opinion}.}

Although our opinion model has rich behavior and provides insights into the effects of group opinions on opinion dynamics, it is important to note that we have not validated our model with real-world opinion data. Indeed, such validation efforts are notoriously {difficult in the study of opinion dynamics~\cite{mas2019,bak2021}, but it can be possible when appropriate data is available \cite{iacopini2024}.} {We hope that further studies of opinion dynamics will encourage and guide efforts in data collection, associated data analysis, and model validation.}


\begin{acknowledgements}

JGR acknowledges support from {the National Science Foundation (grant number DMS-2205967).} MAP was funded in part by the National Science Foundation (grant number 1922952) through their program on Algorithms for Threat Detection. We thank Bard Ermentrout, Zachary Kilpatrick, Ekaterina Landgren, Nicholas Landry, James Meiss, and Nancy Rodr\'{\i}guez for helpful comments. We also thank the {anonymous} referees for their useful comments.

\end{acknowledgements}


\appendix


\section{{Hypergraphs and the random-hypergraph model}}
\label{randhypergraphs}

As {we} stated in Sec.~\ref{twoB}, it is convenient to use hypergraphs to describe our networks, which consist of nodes and groups of nodes. A hypergraph is a generalization of a graph that includes both ordinary edges (i.e., dyadic adjacencies) and hyperedges with more than two nodes (i.e., polyadic adjacencies)~{\cite{newman2018networks,bick2023}}. Following standard convention, we {use the term ``hyperedge" for any of these adjacencies.} Mathematically, a hypergraph {$\mathcal{H}_\mathcal{G} = (\mathcal{V},\mathcal{E})$} consists of a set {$\mathcal{V}$} of nodes and a set $\mathcal{E}$ of hyperedges. Each hyperedge is a nonempty subset of {$\mathcal{V}$}; the number of nodes in this subset is the ``size'' of the hyperedge. In an ordinary graph, each node {$i \in \mathcal{V}$} has an associated degree $k_{\, i}$, which indicates the number of edges that are attached (i.e., ``incident'') to it. {The ``hyperdegree"} of node $i$ is the vector {${\bf k}_i = [k_{\, i}^{(2)},k_{\, i}^{(3)}, \ldots, k_{\, i}^{(L)}]$, where $L$} is the size of its largest hyperedge and the {$l$th-order degree $k_i^{(l)}$ is the number of size-$l$} hyperedges that are incident to node $i$. Each hypergraph has a hyperdegree distribution $\mathcal{P}({\bf k})$, which encodes the probabilities that a uniformly-randomly-chosen node has hyperdegree ${\bf k}$ for each ${\bf k}$.

Just as one can describe a hypergraph using a bipartite network~\cite{newman2018networks}, it is also possible to formulate our {opinion} model by considering dynamics on an ordinary graph with two types of nodes. In this formulation, the agents constitute one type of node and the groups constitute a second type of node. {An agent can have adjacencies both with groups and with other agents, and a group is adjacent to each agent that participates in it.} We view the group language {as} much more natural {than bipartite-network language,} just as the language of hypergraphs and simplicial complexes is natural for studying polyadic interactions \cite{bick2023,torres2021}.

We now describe the particular random-hypergraph model that we employ. {Consider} a set of nodes $i \in \{1,2,3,\ldots, N\}$ with {hyperdegrees} ${\bf k}_i$, {where $k_i^{(l)}$ is the number of loose ends (i.e., ``stubs'') {of} size-$l$} hyperedges that are attached to $i$. We form a {size-$l$ hyperedge by uniformly randomly selecting $l$ stubs} for the hyperedge. We repeat this stub-selection process until all stubs are assigned to a hyperedge. {If $\langle k^{(l)}\rangle$ is finite, the probability that there is a size-$l$ hyperedge that connects nodes $\{i_1,i_2,\ldots, i_l\}$ in the {limit} $N\rightarrow \infty$ is} 
\begin{equation} \label{general_hyperedge_probability}
    {f_l({\bf k}_{i_1}, {\bf k}_{i_2}, \ldots, {\bf k}_{i_l}) = \frac{(l - 1)!\,k^{(l)}_{i_1}k^{(l)}_{i_2}\times \cdots \times \, k^{(l)}_{i_l}}{\left(N\langle k^{(l)}\rangle\right)^{l - 1}}\, . }
\end{equation}
{This expression is a generalization of an associated expression for ordinary configuration-model graphs \cite{fosdick2018configuring,newman2018networks}.}

The above random-hypergraph model is a special case of the stub-labeled hypergraph configuration model in \cite{chodrow2020configuration} in which we allow only hyperedges of sizes 2 and 3. 
{Constructing a hypergraph with a configuration model yields} a small number of self-hyperedges {(i.e., hyperedges in which a single node participates {two} or more times)} and multi-hyperedges {(i.e., {redundant hyperedges, which are hyperedges that occur two or more times})} (See Sec.~12.1 of Ref.~\cite{newman2018networks} for a relevant discussion in the context of graphs.) {Because we assume that nodes do not influence themselves, we remove these self-hyperedges when constructing a hypergraph.} For example, {{{in a calculation} of means using {a single sample of} 20 instantiations} of a 2500-node hypergraph with $\gamma = 4$ and $\langle k \rangle = \langle q \rangle = 20$, we {remove} approximately $0.05$\% of the edges and approximately {$0.15$\%} of the triangles.} However, we retain multi-hyperedges. {From the same sample of 20 instantiations of a 2500-node hypergraph with $\gamma = 4$ and $\langle k \rangle = \langle q \rangle = 20$, we find that about  $0.33\%$ of all edges and about $0.0003\%$ of all triangles are multi-hyperedges.}
 
 Our random-hypergraph model is reminiscent of the closely related random-graph models with clustering that were proposed independently by Newman \cite{newman2009} and Miller \cite{miller2009}. In a Newman--Miller model, nodes have specified dyadic and {triadic (i.e., triangle)} degree sequences, which one uses independently in a stub-matching procedure. {However, a ``triangle" in the Newman--Miller model corresponds to three dyadic interactions, rather than to a single size-3 hyperedge.}


\section{{Detailed derivation of} the mean-field approximation {\eqref{meandyn}}}
\label{meanfield_app}

{In this appendix, we give} {a detailed} derivation of our mean-field {equations}~\eqref{meandyn}. As {we} discussed in Sec.~\ref{three}, {we consider} the time evolution of the three order parameters $V^t$, $U^t$, and $Y^t$. 

{We begin by expressing $V^t$ and $U^t$} in terms of $x_{\bf k}^t$, which is the fraction of nodes at time $t$ with hyperdegree ${\bf k} = [k,q]$ that have opinion $1$.  
{The total numbers of edges is $N \langle k \rangle/2$, and the total number of triangles is $N \langle q \rangle/3$. The expected fraction of opinion-1 nodes in a uniformly randomly selected edge is thus}
\begin{widetext}
\begin{align}
    V^t = \sum_{{\bf k}}\sum _{{\bf k}'} \frac{N\mathcal{P}({\bf k})N\mathcal{P}({\bf k}')}{2!}\left(\frac{k k'}{N\langle k \rangle}\right)\left(\frac{x^t_{{\bf k}} + x^t_{{\bf k}'}}{2} \right){\Big /}\left(\frac{N\langle k \rangle }{2}\right) = \sum_{{\bf k}}\frac{k \mathcal{P}({\bf k})x^t_{{\bf k}}}{\langle k \rangle}\, ,\label{Vapp}
\end{align}
\end{widetext}
where $NP({\bf k})NP({\bf k}')/2!$ is the expected number of pairs of nodes with hyperdegrees ${\bf k}$ and ${\bf k}'$, the quantity $kk'/(N \langle k \rangle)$ is the expected fraction of these pairs that are connected by an edge [see Eq.~\eqref{general_hyperedge_probability} with {$l = 2$}], and $(x_{\bf k}^t + x_{{\bf k}'}^t)/2$ is the expected fraction of opinion-1 nodes that are attached to an edge that connects uniformly-randomly-selected nodes with hyperdegrees ${\bf k}$ and ${\bf k}'$. {Analogously, the expected fraction of opinion-1 nodes in a uniformly randomly selected triangle is}
\begin{widetext}
\begin{align}
  {U^t = \sum _{{\bf k}}\sum_{{\bf k}'}\sum_{{\bf k}''} \frac{N\mathcal{P}({\bf k})\,N\mathcal{P}({\bf k}')\,N\mathcal{P}({\bf k}'')\,}{3!} \left(\frac{2q\,q'\,q''}{(N\langle q \rangle)^2}\right)\left(\frac{x_{{\bf k}}^t+x_{{\bf k}'}^t+x_{{\bf k}''}^t}{3}\right) \Big / \left(\frac{N\langle q \rangle }{3}\right)  
  = \underset{{\bf k}}{\sum} \frac{q\mathcal{P} ({\bf k})x_{{\bf k}}^{t}}{\langle q  \rangle}\,} .\label{Uapp}
\end{align}
\end{widetext}

Because we generate hypergraphs using a configuration model, the probability that there is a hyperedge that connects a group of nodes depends only on the hyperdegrees of those nodes. 
We can thus use a hyperdegree-based compartmental model to obtain mean-field equations. Consider the probability [see Eq.~\eqref{node_model}] that a node $i$ with hyperdegree ${\bf k} = [k,q]$ has opinion $1$ at time $t + 1$. 
Assuming that all nodes with the same hyperdegree behave {in the same way} {(i.e., the probability that node $j$ has opinion 1 is $x_{\bf k}^t$ for all {nodes} $j$ with hyperdegree ${\bf k}_j={\bf k}$)}, we approximate {the normalized number $\bar{x}_i^{\, t}$ of neighbors of node $i$ that have opinion $1$ [see Eq.~(\ref{xbar})] by}
\begin{align}
    \bar{x}_i^{{\, t}} &= \frac{1}{\langle k \rangle}\sum _{j = 1}^NA_{ij}x_j^{t}   \notag\\
    &\approx \frac{1}{\langle k \rangle}\sum _{{\bf k}'}N\mathcal{P}(k',q')\left(\frac{k k'}{N\langle k \rangle}\right)x_{{\bf k}'}^t   \notag \\
    &= \frac{k}{\langle k\rangle}\sum_{{\bf k}'}\frac{k' \mathcal{P}(k',q')x^t_{{\bf k}'}}{\langle k \rangle} \notag \\
    &= kV^t/\langle k\rangle\, , \label{xbarapproxapp}
\end{align}
where the approximation in the second line replaces the number of opinion-1 neighbors of node $i$ with its expected value. The term $\bar{y}_i^{\, t}$, which is the normalized number of triangles that are attached to node $i$ {and} have opinion $1$ [see Eq. (\ref{zbar})], is approximately
\begin{align}
    \bar{y}_i^{{\, t}} \approx q Y^t/\langle q\rangle  \label{ybarapproxapp}
\end{align}
because node $i$ is attached to $q$ triangles and $Y^t$ is the expected fraction of triangles that have opinion 1. We insert the approximations (\ref{xbarapproxapp}) and (\ref{ybarapproxapp}) into Eqs.~(\ref{node_model})--(\ref{node_prob}) to obtain
\begin{align}
    x_{{\bf k}}^{t+1} &= \frac{1}{N\mathcal{P({\bf k})}}\sum _{{\bf k}_i = {\bf k}} {\mathbb E}[x_{i}^{t + 1}]  \notag \\
    &= \frac{1}{N\mathcal{P}({\bf k})}\sum _{{\bf k}_i={\bf k}} f\left(a\bar{x}_i^{\, t} + b\bar{y}_i^{\, t} \right)  \notag \\
    &\approx \frac{1}{N\mathcal{P({\bf k})}}\sum _{{\bf k}_i={\bf k}} f\left(\frac{ak}{\langle k \rangle}V^t + \frac{bq}{\langle q \rangle}Y^t \right) \notag \\
    &= f\left(\frac{ak}{\langle k \rangle}V^t + \frac{bq}{\langle q \rangle}Y^t \right)\, ,\label{xkdynapp}
\end{align}
where ${\mathbb E}[\cdot]$ denotes the expectation. The time evolution of the expected fraction $Y^t$ of triangles with opinion 1 satisfies
\begin{align}
    Y^{t + 1} &= {\mathbb E}\left[ \frac{1}{S}\sum_{j = 1}^Sy_j^{t+1}\right] = \frac{1}{S}\sum_{j=1}^S {\mathbb E}\left[ y_j^{t + 1}\right]  \notag \\
    &= \frac{1}{S}\sum_{j = 1}^S\left \{{\mathbb E}[y_j^{t + 1}|y_j^t = 1]P(y_j^t = 1)\right. \notag \\
    &\qquad \qquad \,\, + \left. {\mathbb E}[y_j^{t + 1}|y_j^t = 0]P(y_j^t = 0) \right \}\, .
\end{align}
Making the mean-field assumption that all triangles behave {in the same way} (i.e., $y_j^t = y^t$ and $\bar{z}^{{\, t}} = \bar{z}_j^{{\, t}}$ for all $j$) yields 
\begin{align}
    Y^{t + 1} &= {\mathbb E}[y^{t + 1}|y^t = 1]P(y^t = 1) \notag  \\
    &\quad + {\mathbb E}[y^{t + 1}|y^t = 0]P(y^t = 0)\,.
\end{align}
We then use Eq.~\eqref{edge_prob} for the expected values and the relation $P(y^t = 1) = Y^t$ to obtain
\begin{align} \label{Ydymwithzapp}
    Y^{t + 1} = f(c \bar{z}^{\, t} + d)Y^t +f(c \bar{z}^{\, t})(1 - Y^t) \,.
\end{align}
Finally, similarly to our approximation of {$\bar{y}_i^{\, t}$ in Eq.~(\ref{ybarapprox}), we approximate $\bar{z}^{\, t}$ (i.e., the fraction of nodes with opinion 1 in a triangle that we select uniformly at random) by
\begin{align} 
    \bar{z}^{{\, t}}&\approx U^t \, , \label{zbarapproxapp}
\end{align}
which is the expected fraction of {opinion-1 nodes at time $t$ in} a triangle that we select uniformly at random.} Substituting Eq.~(\ref{zbarapproxapp}) into Eq.~(\ref{Ydymwithzapp}) yields
\begin{equation}
    Y^{\, t+1} = Y^tf\left(cU^t + d \right) + (1 - Y^t)f(cU^t)\, .\label{Yequapp}
\end{equation}
Inserting Eq.~(\ref{xkdynapp}) into Eqs.~(\ref{Vapp})--(\ref{Uapp}) {yields the closed map~\eqref{meandyn} for the time evolution of the three order parameters:}
\begin{align} \label{meandynapp}
\begin{split}
    V^{\, t + 1} &= \sum_{k}\sum_{q} \frac{k   \mathcal{P} (k,q)}{\langle k \rangle} f\left( \frac{a\,k}{\langle k \rangle}V^t + \frac{b\,q}{\langle q \rangle}Y^t \right)\, ,\\
    U^{\, t + 1} &= \sum_{k}\sum_{q} \frac{q  \mathcal{P} (k,q)}{\langle q \rangle} f\left( \frac{a\,k}{\langle k \rangle}V^t + \frac{b\,q}{\langle q \rangle}Y^t \right)\, ,\\
    Y^{\, t + 1} &= Y^tf\left(cU^t + d \right) + (1 - Y^t)f(cU^t)\, .
\end{split}
\end{align}


\section{{Selection of initial conditions}}\label{ICS}

In this appendix, we discuss how we select initial conditions for our simulations of the stochastic opinion model~\eqref{node_model}--\eqref{edge_model}, the deterministic mean-field approximation~\eqref{meandyn}, and the stochastically perturbed mean-field approximation~\eqref{dynsto}. In most of our simulations, we perform parameter sweeps using {$W^2$} evenly spaced initial conditions in the unit square. For each initial condition, we perform a single simulation of the stochastic opinion {model~\eqref{node_model}--\eqref{edge_model}, the deterministic mean-field map~\eqref{meandyn}, or the stochastically perturbed mean-field map~\eqref{dynsto}.}
 
For the stochastic opinion model~\eqref{node_model}--\eqref{edge_model}, we use the {$W^2$} evenly spaced probabilities $(u_1,u_2)$ (which we described in Sec.~\ref{our-model}) in the set {$\{(i/({W} - 1),\,j/({W} - 1))|\,\, i,j  \in \{0, 1, \ldots , {W} - 1\}\}$} for our parameter sweeps. {For the mean-field equations~\eqref{meandyn} and \eqref{dynsto}, we need to select initial values for the three order parameters $(V^0,U^0,Y^0)$.} {As we described in Sec.~\ref{three}, we employ initial conditions with $V^0 = U^0$ so that our initial conditions resemble our initial conditions for the stochastic opinion model as closely as possible.} Therefore, we use the {$W^2$} evenly spaced initial conditions $(V^0,U^0,Y^0)$ in the set {$\{(i/({W} - 1)),\,i/({W} - 1)),\,j/({W} - 1)))|\,\, i,j \in \{0, 1, \ldots, {W} - 1\}\}$} for our parameter sweeps.


\bibliography{Hyperedge_Opinion-v16}

\providecommand{\noopsort}[1]{}\providecommand{\singleletter}[1]{#1}%
\begin{thebibliography}{81}%
\makeatletter
\providecommand \@ifxundefined [1]{%
 \@ifx{#1\undefined}
}%
\providecommand \@ifnum [1]{%
 \ifnum #1\expandafter \@firstoftwo
 \else \expandafter \@secondoftwo
 \fi
}%
\providecommand \@ifx [1]{%
 \ifx #1\expandafter \@firstoftwo
 \else \expandafter \@secondoftwo
 \fi
}%
\providecommand \natexlab [1]{#1}%
\providecommand \enquote  [1]{``#1''}%
\providecommand \bibnamefont  [1]{#1}%
\providecommand \bibfnamefont [1]{#1}%
\providecommand \citenamefont [1]{#1}%
\providecommand \href@noop [0]{\@secondoftwo}%
\providecommand \href [0]{\begingroup \@sanitize@url \@href}%
\providecommand \@href[1]{\@@startlink{#1}\@@href}%
\providecommand \@@href[1]{\endgroup#1\@@endlink}%
\providecommand \@sanitize@url [0]{\catcode `\\12\catcode `\$12\catcode `\&12\catcode `\#12\catcode `\^12\catcode `\_12\catcode `\%12\relax}%
\providecommand \@@startlink[1]{}%
\providecommand \@@endlink[0]{}%
\providecommand \url  [0]{\begingroup\@sanitize@url \@url }%
\providecommand \@url [1]{\endgroup\@href {#1}{\urlprefix }}%
\providecommand \urlprefix  [0]{URL }%
\providecommand \Eprint [0]{\href }%
\providecommand \doibase [0]{https://doi.org/}%
\providecommand \selectlanguage [0]{\@gobble}%
\providecommand \bibinfo  [0]{\@secondoftwo}%
\providecommand \bibfield  [0]{\@secondoftwo}%
\providecommand \translation [1]{[#1]}%
\providecommand \BibitemOpen [0]{}%
\providecommand \bibitemStop [0]{}%
\providecommand \bibitemNoStop [0]{.\EOS\space}%
\providecommand \EOS [0]{\spacefactor3000\relax}%
\providecommand \BibitemShut  [1]{\csname bibitem#1\endcsname}%
\let\auto@bib@innerbib\@empty
\bibitem [{\citenamefont {Kozitsin}(2023)}]{kozitsin2023opinion}%
  \BibitemOpen
  \bibfield  {author} {\bibinfo {author} {\bibfnamefont {I.~V.}\ \bibnamefont {Kozitsin}},\ }\bibfield  {title} {\bibinfo {title} {Opinion dynamics of online social network users: {A} micro-level analysis},\ }\href@noop {} {\bibfield  {journal} {\bibinfo  {journal} {The Journal of Mathematical Sociology}\ }\textbf {\bibinfo {volume} {47}},\ \bibinfo {pages} {1} (\bibinfo {year} {2023})}\BibitemShut {NoStop}%
\bibitem [{\citenamefont {Zha}\ \emph {et~al.}(2020)\citenamefont {Zha}, \citenamefont {Kou}, \citenamefont {Zhang}, \citenamefont {Liang}, \citenamefont {Chen}, \citenamefont {Li},\ and\ \citenamefont {Dong}}]{zha2020opinion}%
  \BibitemOpen
  \bibfield  {author} {\bibinfo {author} {\bibfnamefont {Q.}~\bibnamefont {Zha}}, \bibinfo {author} {\bibfnamefont {G.}~\bibnamefont {Kou}}, \bibinfo {author} {\bibfnamefont {H.}~\bibnamefont {Zhang}}, \bibinfo {author} {\bibfnamefont {H.}~\bibnamefont {Liang}}, \bibinfo {author} {\bibfnamefont {X.}~\bibnamefont {Chen}}, \bibinfo {author} {\bibfnamefont {C.-C.}\ \bibnamefont {Li}},\ and\ \bibinfo {author} {\bibfnamefont {Y.}~\bibnamefont {Dong}},\ }\bibfield  {title} {\bibinfo {title} {Opinion dynamics in finance and business: {A} literature review and research opportunities},\ }\href@noop {} {\bibfield  {journal} {\bibinfo  {journal} {Financial Innovation}\ }\textbf {\bibinfo {volume} {6}},\ \bibinfo {pages} {44} (\bibinfo {year} {2020})}\BibitemShut {NoStop}%
\bibitem [{\citenamefont {Fern{\'a}ndez-Gracia}\ \emph {et~al.}(2014)\citenamefont {Fern{\'a}ndez-Gracia}, \citenamefont {Suchecki}, \citenamefont {Ramasco}, \citenamefont {San~Miguel},\ and\ \citenamefont {Egu{\'\i}luz}}]{fernandez2014voter}%
  \BibitemOpen
  \bibfield  {author} {\bibinfo {author} {\bibfnamefont {J.}~\bibnamefont {Fern{\'a}ndez-Gracia}}, \bibinfo {author} {\bibfnamefont {K.}~\bibnamefont {Suchecki}}, \bibinfo {author} {\bibfnamefont {J.~J.}\ \bibnamefont {Ramasco}}, \bibinfo {author} {\bibfnamefont {M.}~\bibnamefont {San~Miguel}},\ and\ \bibinfo {author} {\bibfnamefont {V.~M.}\ \bibnamefont {Egu{\'\i}luz}},\ }\bibfield  {title} {\bibinfo {title} {Is the voter model a model for voters?},\ }\href@noop {} {\bibfield  {journal} {\bibinfo  {journal} {Physical Review Letters}\ }\textbf {\bibinfo {volume} {112}},\ \bibinfo {pages} {158701} (\bibinfo {year} {2014})}\BibitemShut {NoStop}%
\bibitem [{\citenamefont {Siegel}(2009)}]{siegel2009social}%
  \BibitemOpen
  \bibfield  {author} {\bibinfo {author} {\bibfnamefont {D.~A.}\ \bibnamefont {Siegel}},\ }\bibfield  {title} {\bibinfo {title} {Social networks and collective action},\ }\href@noop {} {\bibfield  {journal} {\bibinfo  {journal} {American Journal of Political Science}\ }\textbf {\bibinfo {volume} {53}},\ \bibinfo {pages} {122} (\bibinfo {year} {2009})}\BibitemShut {NoStop}%
\bibitem [{\citenamefont {Horstmeyer}\ and\ \citenamefont {Kuehn}(2020)}]{horstmeyer2020adaptive}%
  \BibitemOpen
  \bibfield  {author} {\bibinfo {author} {\bibfnamefont {L.}~\bibnamefont {Horstmeyer}}\ and\ \bibinfo {author} {\bibfnamefont {C.}~\bibnamefont {Kuehn}},\ }\bibfield  {title} {\bibinfo {title} {Adaptive voter model on simplicial complexes},\ }\href@noop {} {\bibfield  {journal} {\bibinfo  {journal} {Physical Review E}\ }\textbf {\bibinfo {volume} {101}},\ \bibinfo {pages} {022305} (\bibinfo {year} {2020})}\BibitemShut {NoStop}%
\bibitem [{\citenamefont {Deffuant}\ \emph {et~al.}(2000)\citenamefont {Deffuant}, \citenamefont {Neau}, \citenamefont {Amblard},\ and\ \citenamefont {Weisbuch}}]{deffuant2000mixing}%
  \BibitemOpen
  \bibfield  {author} {\bibinfo {author} {\bibfnamefont {G.}~\bibnamefont {Deffuant}}, \bibinfo {author} {\bibfnamefont {D.}~\bibnamefont {Neau}}, \bibinfo {author} {\bibfnamefont {F.}~\bibnamefont {Amblard}},\ and\ \bibinfo {author} {\bibfnamefont {G.}~\bibnamefont {Weisbuch}},\ }\bibfield  {title} {\bibinfo {title} {Mixing beliefs among interacting agents},\ }\href@noop {} {\bibfield  {journal} {\bibinfo  {journal} {Advances in Complex Systems}\ }\textbf {\bibinfo {volume} {3}},\ \bibinfo {pages} {87} (\bibinfo {year} {2000})}\BibitemShut {NoStop}%
\bibitem [{\citenamefont {Neuh{\"a}user}\ \emph {et~al.}(2020)\citenamefont {Neuh{\"a}user}, \citenamefont {Mellor},\ and\ \citenamefont {Lambiotte}}]{neuhauser2020multibody}%
  \BibitemOpen
  \bibfield  {author} {\bibinfo {author} {\bibfnamefont {L.}~\bibnamefont {Neuh{\"a}user}}, \bibinfo {author} {\bibfnamefont {A.}~\bibnamefont {Mellor}},\ and\ \bibinfo {author} {\bibfnamefont {R.}~\bibnamefont {Lambiotte}},\ }\bibfield  {title} {\bibinfo {title} {Multibody interactions and nonlinear consensus dynamics on networked systems},\ }\href@noop {} {\bibfield  {journal} {\bibinfo  {journal} {Physical Review E}\ }\textbf {\bibinfo {volume} {101}},\ \bibinfo {pages} {032310} (\bibinfo {year} {2020})}\BibitemShut {NoStop}%
\bibitem [{\citenamefont {Hickok}\ \emph {et~al.}(2022)\citenamefont {Hickok}, \citenamefont {Kureh}, \citenamefont {Brooks}, \citenamefont {Feng},\ and\ \citenamefont {Porter}}]{hickok2022bounded}%
  \BibitemOpen
  \bibfield  {author} {\bibinfo {author} {\bibfnamefont {A.}~\bibnamefont {Hickok}}, \bibinfo {author} {\bibfnamefont {Y.}~\bibnamefont {Kureh}}, \bibinfo {author} {\bibfnamefont {H.~Z.}\ \bibnamefont {Brooks}}, \bibinfo {author} {\bibfnamefont {M.}~\bibnamefont {Feng}},\ and\ \bibinfo {author} {\bibfnamefont {M.~A.}\ \bibnamefont {Porter}},\ }\bibfield  {title} {\bibinfo {title} {A bounded-confidence model of opinion dynamics on hypergraphs},\ }\href@noop {} {\bibfield  {journal} {\bibinfo  {journal} {SIAM Journal on Applied Dynamical Systems}\ }\textbf {\bibinfo {volume} {21}},\ \bibinfo {pages} {1} (\bibinfo {year} {2022})}\BibitemShut {NoStop}%
\bibitem [{\citenamefont {Srivastava}\ \emph {et~al.}(2016)\citenamefont {Srivastava}, \citenamefont {Chelmis},\ and\ \citenamefont {Prasanna}}]{srivastava2016computing}%
  \BibitemOpen
  \bibfield  {author} {\bibinfo {author} {\bibfnamefont {A.}~\bibnamefont {Srivastava}}, \bibinfo {author} {\bibfnamefont {C.}~\bibnamefont {Chelmis}},\ and\ \bibinfo {author} {\bibfnamefont {V.~K.}\ \bibnamefont {Prasanna}},\ }\bibfield  {title} {\bibinfo {title} {Computing competing cascades on signed networks},\ }\href@noop {} {\bibfield  {journal} {\bibinfo  {journal} {Social Network Analysis and Mining}\ }\textbf {\bibinfo {volume} {6}},\ \bibinfo {pages} {82} (\bibinfo {year} {2016})}\BibitemShut {NoStop}%
\bibitem [{\citenamefont {Toccaceli}\ \emph {et~al.}(2020)\citenamefont {Toccaceli}, \citenamefont {Milli},\ and\ \citenamefont {Rossetti}}]{toccaceli2020opinion}%
  \BibitemOpen
  \bibfield  {author} {\bibinfo {author} {\bibfnamefont {C.}~\bibnamefont {Toccaceli}}, \bibinfo {author} {\bibfnamefont {L.}~\bibnamefont {Milli}},\ and\ \bibinfo {author} {\bibfnamefont {G.}~\bibnamefont {Rossetti}},\ }\bibfield  {title} {\bibinfo {title} {Opinion dynamic modeling of fake news perception},\ }in\ \href@noop {} {\emph {\bibinfo {booktitle} {Complex Networks \& Their Applications IX. COMPLEX NETWORKS 2020}}},\ \bibinfo {editor} {edited by\ \bibinfo {editor} {\bibfnamefont {R.~M.}\ \bibnamefont {Benito}}, \bibinfo {editor} {\bibfnamefont {C.}~\bibnamefont {Cherifi}}, \bibinfo {editor} {\bibfnamefont {H.}~\bibnamefont {Cherifi}}, \bibinfo {editor} {\bibfnamefont {E.}~\bibnamefont {Moro}}, \bibinfo {editor} {\bibfnamefont {L.}~\bibnamefont {Rocha}},\ and\ \bibinfo {editor} {\bibfnamefont {M.}~\bibnamefont {Sales-Pardo}}}\ (\bibinfo {organization} {Springer-Verlag},\ \bibinfo {address} {Heidelberg, Germany},\ \bibinfo {year} {2020})\ pp.\ \bibinfo {pages} {370--381}\BibitemShut {NoStop}%
\bibitem [{\citenamefont {Maleti{\'c}}\ and\ \citenamefont {Rajkovi{\'c}}(2014)}]{maletic2014consensus}%
  \BibitemOpen
  \bibfield  {author} {\bibinfo {author} {\bibfnamefont {S.}~\bibnamefont {Maleti{\'c}}}\ and\ \bibinfo {author} {\bibfnamefont {M.}~\bibnamefont {Rajkovi{\'c}}},\ }\bibfield  {title} {\bibinfo {title} {Consensus formation on a simplicial complex of opinions},\ }\href@noop {} {\bibfield  {journal} {\bibinfo  {journal} {Physica A: Statistical Mechanics and its Applications}\ }\textbf {\bibinfo {volume} {397}},\ \bibinfo {pages} {111} (\bibinfo {year} {2014})}\BibitemShut {NoStop}%
\bibitem [{\citenamefont {Sahasrabuddhe}\ \emph {et~al.}(2021)\citenamefont {Sahasrabuddhe}, \citenamefont {Neuh{\"a}user},\ and\ \citenamefont {Lambiotte}}]{sahasrabuddhe2021modelling}%
  \BibitemOpen
  \bibfield  {author} {\bibinfo {author} {\bibfnamefont {R.}~\bibnamefont {Sahasrabuddhe}}, \bibinfo {author} {\bibfnamefont {L.}~\bibnamefont {Neuh{\"a}user}},\ and\ \bibinfo {author} {\bibfnamefont {R.}~\bibnamefont {Lambiotte}},\ }\bibfield  {title} {\bibinfo {title} {Modelling non-linear consensus dynamics on hypergraphs},\ }\href@noop {} {\bibfield  {journal} {\bibinfo  {journal} {Journal of Physics: Complexity}\ }\textbf {\bibinfo {volume} {2}},\ \bibinfo {pages} {025006} (\bibinfo {year} {2021})}\BibitemShut {NoStop}%
\bibitem [{\citenamefont {Baumann}\ \emph {et~al.}(2020)\citenamefont {Baumann}, \citenamefont {Lorenz-Spreen}, \citenamefont {Sokolov},\ and\ \citenamefont {Starnini}}]{baumann2020modeling}%
  \BibitemOpen
  \bibfield  {author} {\bibinfo {author} {\bibfnamefont {F.}~\bibnamefont {Baumann}}, \bibinfo {author} {\bibfnamefont {P.}~\bibnamefont {Lorenz-Spreen}}, \bibinfo {author} {\bibfnamefont {I.~M.}\ \bibnamefont {Sokolov}},\ and\ \bibinfo {author} {\bibfnamefont {M.}~\bibnamefont {Starnini}},\ }\bibfield  {title} {\bibinfo {title} {Modeling echo chambers and polarization dynamics in social networks},\ }\href@noop {} {\bibfield  {journal} {\bibinfo  {journal} {Physical Review Letters}\ }\textbf {\bibinfo {volume} {124}},\ \bibinfo {pages} {048301} (\bibinfo {year} {2020})}\BibitemShut {NoStop}%
\bibitem [{\citenamefont {Brooks}\ \emph {et~al.}(2024)\citenamefont {Brooks}, \citenamefont {Chodrow},\ and\ \citenamefont {Porter}}]{brooks2023}%
  \BibitemOpen
  \bibfield  {author} {\bibinfo {author} {\bibfnamefont {H.~Z.}\ \bibnamefont {Brooks}}, \bibinfo {author} {\bibfnamefont {P.~S.}\ \bibnamefont {Chodrow}},\ and\ \bibinfo {author} {\bibfnamefont {M.~A.}\ \bibnamefont {Porter}},\ }\bibfield  {title} {\bibinfo {title} {Emergence of polarization in a sigmoidal bounded-confidence model of opinion dynamics},\ }\href@noop {} {\bibfield  {journal} {\bibinfo  {journal} {SIAM Journal on Applied Dynamical Systems}\ }\textbf {\bibinfo {volume} {23}},\ \bibinfo {pages} {1442} (\bibinfo {year} {2024})}\BibitemShut {NoStop}%
\bibitem [{\citenamefont {S{\^\i}rbu}\ \emph {et~al.}(2017)\citenamefont {S{\^\i}rbu}, \citenamefont {Loreto}, \citenamefont {Servedio},\ and\ \citenamefont {Tria}}]{sirbu2017opinion}%
  \BibitemOpen
  \bibfield  {author} {\bibinfo {author} {\bibfnamefont {A.}~\bibnamefont {S{\^\i}rbu}}, \bibinfo {author} {\bibfnamefont {V.}~\bibnamefont {Loreto}}, \bibinfo {author} {\bibfnamefont {V.~D.~P.}\ \bibnamefont {Servedio}},\ and\ \bibinfo {author} {\bibfnamefont {F.}~\bibnamefont {Tria}},\ }\bibfield  {title} {\bibinfo {title} {Opinion dynamics: {M}odels, extensions and external effects},\ }in\ \href@noop {} {\emph {\bibinfo {booktitle} {Participatory Sensing, Opinions and Collective Awareness}}},\ \bibinfo {editor} {edited by\ \bibinfo {editor} {\bibfnamefont {V.}~\bibnamefont {Loreto}}, \bibinfo {editor} {\bibfnamefont {M.}~\bibnamefont {Haklay}}, \bibinfo {editor} {\bibfnamefont {A.}~\bibnamefont {Hotho}}, \bibinfo {editor} {\bibfnamefont {V.~D.}\ \bibnamefont {Servedio}}, \bibinfo {editor} {\bibfnamefont {G.}~\bibnamefont {Stumme}}, \bibinfo {editor} {\bibfnamefont {J.}~\bibnamefont {Theunis}},\ and\ \bibinfo {editor} {\bibfnamefont {F.}~\bibnamefont {Tria}}}\ (\bibinfo  {publisher} {Springer International
  Publishing},\ \bibinfo {address} {Cham, Switzerland},\ \bibinfo {year} {2017})\ pp.\ \bibinfo {pages} {363--401}\BibitemShut {NoStop}%
\bibitem [{\citenamefont {Noorazar}\ \emph {et~al.}(2020)\citenamefont {Noorazar}, \citenamefont {Vixie}, \citenamefont {Talebanpour},\ and\ \citenamefont {Hu}}]{noorazar2020classical}%
  \BibitemOpen
  \bibfield  {author} {\bibinfo {author} {\bibfnamefont {H.}~\bibnamefont {Noorazar}}, \bibinfo {author} {\bibfnamefont {K.~R.}\ \bibnamefont {Vixie}}, \bibinfo {author} {\bibfnamefont {A.}~\bibnamefont {Talebanpour}},\ and\ \bibinfo {author} {\bibfnamefont {Y.}~\bibnamefont {Hu}},\ }\bibfield  {title} {\bibinfo {title} {From classical to modern opinion dynamics},\ }\href@noop {} {\bibfield  {journal} {\bibinfo  {journal} {International Journal of Modern Physics C}\ }\textbf {\bibinfo {volume} {31}},\ \bibinfo {pages} {2050101} (\bibinfo {year} {2020})}\BibitemShut {NoStop}%
\bibitem [{\citenamefont {Olsson}\ and\ \citenamefont {Galesic}(2024)}]{olsson2023}%
  \BibitemOpen
  \bibfield  {author} {\bibinfo {author} {\bibfnamefont {H.}~\bibnamefont {Olsson}}\ and\ \bibinfo {author} {\bibfnamefont {M.}~\bibnamefont {Galesic}},\ }\bibfield  {title} {\bibinfo {title} {Analogies for modeling belief dynamics},\ }\href@noop {} {\bibfield  {journal} {\bibinfo  {journal} {Trends in Cognitive Sciences}\ }\textbf {\bibinfo {volume} {28}},\ \bibinfo {pages} {907} (\bibinfo {year} {2024})}\BibitemShut {NoStop}%
\bibitem [{\citenamefont {Xie}\ \emph {et~al.}(2016)\citenamefont {Xie}, \citenamefont {Song},\ and\ \citenamefont {Li}}]{xie2016review}%
  \BibitemOpen
  \bibfield  {author} {\bibinfo {author} {\bibfnamefont {Z.}~\bibnamefont {Xie}}, \bibinfo {author} {\bibfnamefont {X.}~\bibnamefont {Song}},\ and\ \bibinfo {author} {\bibfnamefont {Q.}~\bibnamefont {Li}},\ }\bibfield  {title} {\bibinfo {title} {A review of opinion dynamics},\ }in\ \href@noop {} {\emph {\bibinfo {booktitle} {Theory, Methodology, Tools and Applications for Modeling and Simulation of Complex Systems: 16th Asia Simulation Conference and SCS Autumn Simulation Multi-Conference, AsiaSim/SCS AutumnSim 2016, Beijing, China, October 8--11, 2016, Proceedings, Part IV 16}}}\ (\bibinfo {organization} {Springer-Verlag},\ \bibinfo {address} {Heidelberg, Germany},\ \bibinfo {year} {2016})\ pp.\ \bibinfo {pages} {349--357}\BibitemShut {NoStop}%
\bibitem [{\citenamefont {DeGroot}(1974)}]{degroot1974reaching}%
  \BibitemOpen
  \bibfield  {author} {\bibinfo {author} {\bibfnamefont {M.~H.}\ \bibnamefont {DeGroot}},\ }\bibfield  {title} {\bibinfo {title} {Reaching a consensus},\ }\href@noop {} {\bibfield  {journal} {\bibinfo  {journal} {Journal of the American Statistical Association}\ }\textbf {\bibinfo {volume} {69}},\ \bibinfo {pages} {118} (\bibinfo {year} {1974})}\BibitemShut {NoStop}%
\bibitem [{\citenamefont {Clifford}\ and\ \citenamefont {Sudbury}(1973)}]{clifford1973model}%
  \BibitemOpen
  \bibfield  {author} {\bibinfo {author} {\bibfnamefont {P.}~\bibnamefont {Clifford}}\ and\ \bibinfo {author} {\bibfnamefont {A.}~\bibnamefont {Sudbury}},\ }\bibfield  {title} {\bibinfo {title} {A model for spatial conflict},\ }\href@noop {} {\bibfield  {journal} {\bibinfo  {journal} {Biometrika}\ }\textbf {\bibinfo {volume} {60}},\ \bibinfo {pages} {581} (\bibinfo {year} {1973})}\BibitemShut {NoStop}%
\bibitem [{\citenamefont {Holley}\ and\ \citenamefont {Liggett}(1975)}]{holley1975ergodic}%
  \BibitemOpen
  \bibfield  {author} {\bibinfo {author} {\bibfnamefont {R.~A.}\ \bibnamefont {Holley}}\ and\ \bibinfo {author} {\bibfnamefont {T.~M.}\ \bibnamefont {Liggett}},\ }\bibfield  {title} {\bibinfo {title} {Ergodic theorems for weakly interacting infinite systems and the voter model},\ }\href@noop {} {\bibfield  {journal} {\bibinfo  {journal} {The Annals of Probability}\ }\textbf {\bibinfo {volume} {3}},\ \bibinfo {pages} {643} (\bibinfo {year} {1975})}\BibitemShut {NoStop}%
\bibitem [{\citenamefont {Redner}(2019)}]{redner2019}%
  \BibitemOpen
  \bibfield  {author} {\bibinfo {author} {\bibfnamefont {S.}~\bibnamefont {Redner}},\ }\bibfield  {title} {\bibinfo {title} {Reality-inspired voter models: {A} mini-review},\ }\href@noop {} {\bibfield  {journal} {\bibinfo  {journal} {Comptes Rendus Physique}\ }\textbf {\bibinfo {volume} {20}},\ \bibinfo {pages} {275} (\bibinfo {year} {2019})}\BibitemShut {NoStop}%
\bibitem [{\citenamefont {Galam}(2002)}]{Galam02}%
  \BibitemOpen
  \bibfield  {author} {\bibinfo {author} {\bibfnamefont {S.}~\bibnamefont {Galam}},\ }\bibfield  {title} {\bibinfo {title} {Minority opinion spreading in random geometry},\ }\href@noop {} {\bibfield  {journal} {\bibinfo  {journal} {The European Physical Journal B}\ }\textbf {\bibinfo {volume} {25}},\ \bibinfo {pages} {403} (\bibinfo {year} {2002})}\BibitemShut {NoStop}%
\bibitem [{\citenamefont {Bernardo}\ \emph {et~al.}(2024)\citenamefont {Bernardo}, \citenamefont {Altafini}, \citenamefont {Proskurnikov},\ and\ \citenamefont {Vasca}}]{bernardo2024}%
  \BibitemOpen
  \bibfield  {author} {\bibinfo {author} {\bibfnamefont {C.}~\bibnamefont {Bernardo}}, \bibinfo {author} {\bibfnamefont {C.}~\bibnamefont {Altafini}}, \bibinfo {author} {\bibfnamefont {A.}~\bibnamefont {Proskurnikov}},\ and\ \bibinfo {author} {\bibfnamefont {F.}~\bibnamefont {Vasca}},\ }\bibfield  {title} {\bibinfo {title} {Bounded confidence opinion dynamics: {A} survey},\ }\href@noop {} {\bibfield  {journal} {\bibinfo  {journal} {Automatica}\ }\textbf {\bibinfo {volume} {159}},\ \bibinfo {eid} {111302} (\bibinfo {year} {2024})}\BibitemShut {NoStop}%
\bibitem [{\citenamefont {Breiger}(1974)}]{breiger1974duality}%
  \BibitemOpen
  \bibfield  {author} {\bibinfo {author} {\bibfnamefont {R.~L.}\ \bibnamefont {Breiger}},\ }\bibfield  {title} {\bibinfo {title} {The duality of persons and groups},\ }\href@noop {} {\bibfield  {journal} {\bibinfo  {journal} {Social forces}\ }\textbf {\bibinfo {volume} {53}},\ \bibinfo {pages} {181} (\bibinfo {year} {1974})}\BibitemShut {NoStop}%
\bibitem [{\citenamefont {Burningham}\ and\ \citenamefont {West}(1995)}]{burningham1995individual}%
  \BibitemOpen
  \bibfield  {author} {\bibinfo {author} {\bibfnamefont {C.}~\bibnamefont {Burningham}}\ and\ \bibinfo {author} {\bibfnamefont {M.~A.}\ \bibnamefont {West}},\ }\bibfield  {title} {\bibinfo {title} {Individual, climate, and group interaction processes as predictors of work team innovation},\ }\href@noop {} {\bibfield  {journal} {\bibinfo  {journal} {Small Group Research}\ }\textbf {\bibinfo {volume} {26}},\ \bibinfo {pages} {106} (\bibinfo {year} {1995})}\BibitemShut {NoStop}%
\bibitem [{\citenamefont {Fine}(2014)}]{fine2014hinge}%
  \BibitemOpen
  \bibfield  {author} {\bibinfo {author} {\bibfnamefont {G.~A.}\ \bibnamefont {Fine}},\ }\bibfield  {title} {\bibinfo {title} {The hinge: {C}ivil society, group culture, and the interaction order},\ }\href@noop {} {\bibfield  {journal} {\bibinfo  {journal} {Social Psychology Quarterly}\ }\textbf {\bibinfo {volume} {77}},\ \bibinfo {pages} {5} (\bibinfo {year} {2014})}\BibitemShut {NoStop}%
\bibitem [{\citenamefont {Eilert}\ and\ \citenamefont {Nappier~Cherup}(2020)}]{eilert2020activist}%
  \BibitemOpen
  \bibfield  {author} {\bibinfo {author} {\bibfnamefont {M.}~\bibnamefont {Eilert}}\ and\ \bibinfo {author} {\bibfnamefont {A.}~\bibnamefont {Nappier~Cherup}},\ }\bibfield  {title} {\bibinfo {title} {The activist company: {E}xamining a company's pursuit of societal change through corporate activism using an institutional theoretical lens},\ }\href@noop {} {\bibfield  {journal} {\bibinfo  {journal} {Journal of Public Policy \& Marketing}\ }\textbf {\bibinfo {volume} {39}},\ \bibinfo {pages} {461} (\bibinfo {year} {2020})}\BibitemShut {NoStop}%
\bibitem [{\citenamefont {Black}\ \emph {et~al.}(2016)\citenamefont {Black}, \citenamefont {Owens}, \citenamefont {Wedeking},\ and\ \citenamefont {Wohlfarth}}]{black2016us}%
  \BibitemOpen
  \bibfield  {author} {\bibinfo {author} {\bibfnamefont {R.~C.}\ \bibnamefont {Black}}, \bibinfo {author} {\bibfnamefont {R.~J.}\ \bibnamefont {Owens}}, \bibinfo {author} {\bibfnamefont {J.}~\bibnamefont {Wedeking}},\ and\ \bibinfo {author} {\bibfnamefont {P.~C.}\ \bibnamefont {Wohlfarth}},\ }\href@noop {} {\emph {\bibinfo {title} {US Supreme Court Opinions and Their Audiences}}}\ (\bibinfo  {publisher} {Cambridge University Press},\ \bibinfo {address} {Cambridge, UK},\ \bibinfo {year} {2016})\BibitemShut {NoStop}%
\bibitem [{\citenamefont {Cialdini}\ and\ \citenamefont {Goldstein}(2004)}]{cialdini2004social}%
  \BibitemOpen
  \bibfield  {author} {\bibinfo {author} {\bibfnamefont {R.~B.}\ \bibnamefont {Cialdini}}\ and\ \bibinfo {author} {\bibfnamefont {N.~J.}\ \bibnamefont {Goldstein}},\ }\bibfield  {title} {\bibinfo {title} {Social influence: {C}ompliance and conformity},\ }\href@noop {} {\bibfield  {journal} {\bibinfo  {journal} {Annual Review of Psychology}\ }\textbf {\bibinfo {volume} {55}},\ \bibinfo {pages} {591} (\bibinfo {year} {2004})}\BibitemShut {NoStop}%
\bibitem [{\citenamefont {Griskevicius}\ \emph {et~al.}(2006)\citenamefont {Griskevicius}, \citenamefont {Goldstein}, \citenamefont {Mortensen}, \citenamefont {Cialdini},\ and\ \citenamefont {Kenrick}}]{griskevicius2006going}%
  \BibitemOpen
  \bibfield  {author} {\bibinfo {author} {\bibfnamefont {V.}~\bibnamefont {Griskevicius}}, \bibinfo {author} {\bibfnamefont {N.~J.}\ \bibnamefont {Goldstein}}, \bibinfo {author} {\bibfnamefont {C.~R.}\ \bibnamefont {Mortensen}}, \bibinfo {author} {\bibfnamefont {R.~B.}\ \bibnamefont {Cialdini}},\ and\ \bibinfo {author} {\bibfnamefont {D.~T.}\ \bibnamefont {Kenrick}},\ }\bibfield  {title} {\bibinfo {title} {Going along versus going alone: {W}hen fundamental motives facilitate strategic (non)conformity},\ }\href@noop {} {\bibfield  {journal} {\bibinfo  {journal} {Journal of Personality and Social Psychology}\ }\textbf {\bibinfo {volume} {91}},\ \bibinfo {pages} {281} (\bibinfo {year} {2006})}\BibitemShut {NoStop}%
\bibitem [{\citenamefont {Sepulchre}\ \emph {et~al.}(2018)\citenamefont {Sepulchre}, \citenamefont {Drion},\ and\ \citenamefont {Franci}}]{sepulchre2018excitable}%
  \BibitemOpen
  \bibfield  {author} {\bibinfo {author} {\bibfnamefont {R.}~\bibnamefont {Sepulchre}}, \bibinfo {author} {\bibfnamefont {G.}~\bibnamefont {Drion}},\ and\ \bibinfo {author} {\bibfnamefont {A.}~\bibnamefont {Franci}},\ }\bibfield  {title} {\bibinfo {title} {Excitable behaviors},\ }in\ \href@noop {} {\emph {\bibinfo {booktitle} {Emerging Applications of Control and Systems Theory: A Festschrift in Honor of Mathukumalli Vidyasagar}}}\ (\bibinfo  {publisher} {Springer-Verlag},\ \bibinfo {address} {Heidelberg, Germany},\ \bibinfo {year} {2018})\ pp.\ \bibinfo {pages} {269--280}\BibitemShut {NoStop}%
\bibitem [{\citenamefont {Izhikevich}(2007)}]{izhikevich2007dynamical}%
  \BibitemOpen
  \bibfield  {author} {\bibinfo {author} {\bibfnamefont {E.~M.}\ \bibnamefont {Izhikevich}},\ }\href@noop {} {\emph {\bibinfo {title} {Dynamical Systems in Neuroscience}}}\ (\bibinfo  {publisher} {MIT Press},\ \bibinfo {address} {Cambridge, MA, USA},\ \bibinfo {year} {2007})\BibitemShut {NoStop}%
\bibitem [{\citenamefont {Best}(2006)}]{best2006flavor}%
  \BibitemOpen
  \bibfield  {author} {\bibinfo {author} {\bibfnamefont {J.}~\bibnamefont {Best}},\ }\href@noop {} {\emph {\bibinfo {title} {Flavor of the Month: Why Smart People Fall for Fads}}}\ (\bibinfo  {publisher} {University of California Press},\ \bibinfo {address} {Oakland, CA, USA},\ \bibinfo {year} {2006})\BibitemShut {NoStop}%
\bibitem [{\citenamefont {Aguirre}\ \emph {et~al.}(1988)\citenamefont {Aguirre}, \citenamefont {Quarantelli},\ and\ \citenamefont {Mendoza}}]{aguirre1988collective}%
  \BibitemOpen
  \bibfield  {author} {\bibinfo {author} {\bibfnamefont {B.~E.}\ \bibnamefont {Aguirre}}, \bibinfo {author} {\bibfnamefont {E.~L.}\ \bibnamefont {Quarantelli}},\ and\ \bibinfo {author} {\bibfnamefont {J.~L.}\ \bibnamefont {Mendoza}},\ }\bibfield  {title} {\bibinfo {title} {The collective behavior of fads: {T}he characteristics, effects, and career of streaking},\ }\href@noop {} {\bibfield  {journal} {\bibinfo  {journal} {American Sociological Review}\ }\textbf {\bibinfo {volume} {53}},\ \bibinfo {pages} {569} (\bibinfo {year} {1988})}\BibitemShut {NoStop}%
\bibitem [{\citenamefont {Battiston}\ \emph {et~al.}(2020)\citenamefont {Battiston}, \citenamefont {Cencetti}, \citenamefont {Iacopini}, \citenamefont {Latora}, \citenamefont {Lucas}, \citenamefont {Patania}, \citenamefont {Young},\ and\ \citenamefont {Petri}}]{battiston2020networks}%
  \BibitemOpen
  \bibfield  {author} {\bibinfo {author} {\bibfnamefont {F.}~\bibnamefont {Battiston}}, \bibinfo {author} {\bibfnamefont {G.}~\bibnamefont {Cencetti}}, \bibinfo {author} {\bibfnamefont {I.}~\bibnamefont {Iacopini}}, \bibinfo {author} {\bibfnamefont {V.}~\bibnamefont {Latora}}, \bibinfo {author} {\bibfnamefont {M.}~\bibnamefont {Lucas}}, \bibinfo {author} {\bibfnamefont {A.}~\bibnamefont {Patania}}, \bibinfo {author} {\bibfnamefont {J.-G.}\ \bibnamefont {Young}},\ and\ \bibinfo {author} {\bibfnamefont {G.}~\bibnamefont {Petri}},\ }\bibfield  {title} {\bibinfo {title} {Networks beyond pairwise interactions: {S}tructure and dynamics},\ }\href@noop {} {\bibfield  {journal} {\bibinfo  {journal} {Physics Reports}\ }\textbf {\bibinfo {volume} {874}},\ \bibinfo {pages} {1} (\bibinfo {year} {2020})}\BibitemShut {NoStop}%
\bibitem [{\citenamefont {Battiston}\ \emph {et~al.}(2021)\citenamefont {Battiston}, \citenamefont {Amico}, \citenamefont {Barrat}, \citenamefont {Bianconi}, \citenamefont {Ferraz~de Arruda}, \citenamefont {Franceschiello}, \citenamefont {Iacopini}, \citenamefont {K\'{e}fi}, \citenamefont {Latora}, \citenamefont {Moreno}, \citenamefont {Murray}, \citenamefont {Peixoto}, \citenamefont {Vaccarino},\ and\ \citenamefont {Petri}}]{battiston2021}%
  \BibitemOpen
  \bibfield  {author} {\bibinfo {author} {\bibfnamefont {F.}~\bibnamefont {Battiston}}, \bibinfo {author} {\bibfnamefont {E.}~\bibnamefont {Amico}}, \bibinfo {author} {\bibfnamefont {A.}~\bibnamefont {Barrat}}, \bibinfo {author} {\bibfnamefont {G.}~\bibnamefont {Bianconi}}, \bibinfo {author} {\bibfnamefont {G.}~\bibnamefont {Ferraz~de Arruda}}, \bibinfo {author} {\bibfnamefont {B.}~\bibnamefont {Franceschiello}}, \bibinfo {author} {\bibfnamefont {I.}~\bibnamefont {Iacopini}}, \bibinfo {author} {\bibfnamefont {S.}~\bibnamefont {K\'{e}fi}}, \bibinfo {author} {\bibfnamefont {V.}~\bibnamefont {Latora}}, \bibinfo {author} {\bibfnamefont {Y.}~\bibnamefont {Moreno}}, \bibinfo {author} {\bibfnamefont {M.~M.}\ \bibnamefont {Murray}}, \bibinfo {author} {\bibfnamefont {T.~P.}\ \bibnamefont {Peixoto}}, \bibinfo {author} {\bibfnamefont {F.}~\bibnamefont {Vaccarino}},\ and\ \bibinfo {author} {\bibfnamefont {G.}~\bibnamefont {Petri}},\ }\bibfield  {title} {\bibinfo {title} {The physics of higher-order interactions in
  complex systems},\ }\href@noop {} {\bibfield  {journal} {\bibinfo  {journal} {Nature Physics}\ }\textbf {\bibinfo {volume} {17}},\ \bibinfo {pages} {1093} (\bibinfo {year} {2021})}\BibitemShut {NoStop}%
\bibitem [{\citenamefont {Bick}\ \emph {et~al.}(2023)\citenamefont {Bick}, \citenamefont {Gross}, \citenamefont {Harrington},\ and\ \citenamefont {Schaub}}]{bick2023}%
  \BibitemOpen
  \bibfield  {author} {\bibinfo {author} {\bibfnamefont {C.}~\bibnamefont {Bick}}, \bibinfo {author} {\bibfnamefont {E.}~\bibnamefont {Gross}}, \bibinfo {author} {\bibfnamefont {H.~A.}\ \bibnamefont {Harrington}},\ and\ \bibinfo {author} {\bibfnamefont {M.~T.}\ \bibnamefont {Schaub}},\ }\bibfield  {title} {\bibinfo {title} {What are higher-order networks?},\ }\href@noop {} {\bibfield  {journal} {\bibinfo  {journal} {SIAM Review}\ }\textbf {\bibinfo {volume} {65}},\ \bibinfo {pages} {686} (\bibinfo {year} {2023})}\BibitemShut {NoStop}%
\bibitem [{\citenamefont {Gao}\ \emph {et~al.}(2023)\citenamefont {Gao}, \citenamefont {Ghosh}, \citenamefont {Harrington}, \citenamefont {Restrepo},\ and\ \citenamefont {Taylor}}]{gao2023dynamics}%
  \BibitemOpen
  \bibfield  {author} {\bibinfo {author} {\bibfnamefont {Z.}~\bibnamefont {Gao}}, \bibinfo {author} {\bibfnamefont {D.}~\bibnamefont {Ghosh}}, \bibinfo {author} {\bibfnamefont {H.~A.}\ \bibnamefont {Harrington}}, \bibinfo {author} {\bibfnamefont {J.~G.}\ \bibnamefont {Restrepo}},\ and\ \bibinfo {author} {\bibfnamefont {D.}~\bibnamefont {Taylor}},\ }\bibfield  {title} {\bibinfo {title} {Dynamics on networks with higher-order interactions},\ }\href@noop {} {\bibfield  {journal} {\bibinfo  {journal} {Chaos: An Interdisciplinary Journal of Nonlinear Science}\ }\textbf {\bibinfo {volume} {33}},\ \bibinfo {pages} {040401} (\bibinfo {year} {2023})}\BibitemShut {NoStop}%
\bibitem [{\citenamefont {Atkin}(1977)}]{atkin1977combinatorial}%
  \BibitemOpen
  \bibfield  {author} {\bibinfo {author} {\bibfnamefont {R.~H.}\ \bibnamefont {Atkin}},\ }\href@noop {} {\emph {\bibinfo {title} {Combinatorial Connectivities in Social Systems: An Application of Simplicial Complex Structures to the Study of Large Organizations}}}\ (\bibinfo  {publisher} {Springer-Verlag},\ \bibinfo {address} {Heidelberg, Germany},\ \bibinfo {year} {1977})\BibitemShut {NoStop}%
\bibitem [{\citenamefont {Abrams}(1983)}]{abrams1983arguments}%
  \BibitemOpen
  \bibfield  {author} {\bibinfo {author} {\bibfnamefont {P.~A.}\ \bibnamefont {Abrams}},\ }\bibfield  {title} {\bibinfo {title} {Arguments in favor of higher order interactions},\ }\href@noop {} {\bibfield  {journal} {\bibinfo  {journal} {The American Naturalist}\ }\textbf {\bibinfo {volume} {121}},\ \bibinfo {pages} {887} (\bibinfo {year} {1983})}\BibitemShut {NoStop}%
\bibitem [{\citenamefont {Wilbur}\ and\ \citenamefont {Fauth}(1990)}]{wilbur1990experimental}%
  \BibitemOpen
  \bibfield  {author} {\bibinfo {author} {\bibfnamefont {H.~M.}\ \bibnamefont {Wilbur}}\ and\ \bibinfo {author} {\bibfnamefont {J.~E.}\ \bibnamefont {Fauth}},\ }\bibfield  {title} {\bibinfo {title} {Experimental aquatic food webs: {I}nteractions between two predators and two prey},\ }\href@noop {} {\bibfield  {journal} {\bibinfo  {journal} {The American Naturalist}\ }\textbf {\bibinfo {volume} {135}},\ \bibinfo {pages} {176} (\bibinfo {year} {1990})}\BibitemShut {NoStop}%
\bibitem [{\citenamefont {Schaffer}\ and\ \citenamefont {Liddell}(1984)}]{schaffer1984adult}%
  \BibitemOpen
  \bibfield  {author} {\bibinfo {author} {\bibfnamefont {H.}~\bibnamefont {Schaffer}}\ and\ \bibinfo {author} {\bibfnamefont {C.}~\bibnamefont {Liddell}},\ }\bibfield  {title} {\bibinfo {title} {Adult--child interaction under dyadic and polyadic conditions},\ }\href@noop {} {\bibfield  {journal} {\bibinfo  {journal} {British Journal of Developmental Psychology}\ }\textbf {\bibinfo {volume} {2}},\ \bibinfo {pages} {33} (\bibinfo {year} {1984})}\BibitemShut {NoStop}%
\bibitem [{\citenamefont {Iacopini}\ \emph {et~al.}(2019)\citenamefont {Iacopini}, \citenamefont {Petri}, \citenamefont {Barrat},\ and\ \citenamefont {Latora}}]{iacopini2019simplicial}%
  \BibitemOpen
  \bibfield  {author} {\bibinfo {author} {\bibfnamefont {I.}~\bibnamefont {Iacopini}}, \bibinfo {author} {\bibfnamefont {G.}~\bibnamefont {Petri}}, \bibinfo {author} {\bibfnamefont {A.}~\bibnamefont {Barrat}},\ and\ \bibinfo {author} {\bibfnamefont {V.}~\bibnamefont {Latora}},\ }\bibfield  {title} {\bibinfo {title} {Simplicial models of social contagion},\ }\href@noop {} {\bibfield  {journal} {\bibinfo  {journal} {Nature Communications}\ }\textbf {\bibinfo {volume} {10}},\ \bibinfo {pages} {2485} (\bibinfo {year} {2019})}\BibitemShut {NoStop}%
\bibitem [{\citenamefont {Neuh{\"a}user}\ \emph {et~al.}(2021)\citenamefont {Neuh{\"a}user}, \citenamefont {Schaub}, \citenamefont {Mellor},\ and\ \citenamefont {Lambiotte}}]{neuhauser2021opinion}%
  \BibitemOpen
  \bibfield  {author} {\bibinfo {author} {\bibfnamefont {L.}~\bibnamefont {Neuh{\"a}user}}, \bibinfo {author} {\bibfnamefont {M.~T.}\ \bibnamefont {Schaub}}, \bibinfo {author} {\bibfnamefont {A.}~\bibnamefont {Mellor}},\ and\ \bibinfo {author} {\bibfnamefont {R.}~\bibnamefont {Lambiotte}},\ }\bibfield  {title} {\bibinfo {title} {Opinion dynamics with multi-body interactions},\ }in\ \href@noop {} {\emph {\bibinfo {booktitle} {International Conference on Network Games, Control and Optimization. NETGCOOP 2021.}}},\ \bibinfo {editor} {edited by\ \bibinfo {editor} {\bibfnamefont {S.}~\bibnamefont {Lasaulce}}, \bibinfo {editor} {\bibfnamefont {P.}~\bibnamefont {Mertikopoulos}},\ and\ \bibinfo {editor} {\bibfnamefont {A.}~\bibnamefont {Orda}}}\ (\bibinfo {organization} {Springer International Publishing},\ \bibinfo {address} {Cham, Switzerland},\ \bibinfo {year} {2021})\ pp.\ \bibinfo {pages} {261--271}\BibitemShut {NoStop}%
\bibitem [{\citenamefont {Cencetti}\ \emph {et~al.}(2021)\citenamefont {Cencetti}, \citenamefont {Battiston}, \citenamefont {Lepri},\ and\ \citenamefont {Karsai}}]{cencetti2021temporal}%
  \BibitemOpen
  \bibfield  {author} {\bibinfo {author} {\bibfnamefont {G.}~\bibnamefont {Cencetti}}, \bibinfo {author} {\bibfnamefont {F.}~\bibnamefont {Battiston}}, \bibinfo {author} {\bibfnamefont {B.}~\bibnamefont {Lepri}},\ and\ \bibinfo {author} {\bibfnamefont {M.}~\bibnamefont {Karsai}},\ }\bibfield  {title} {\bibinfo {title} {Temporal properties of higher-order interactions in social networks},\ }\href@noop {} {\bibfield  {journal} {\bibinfo  {journal} {Scientific Reports}\ }\textbf {\bibinfo {volume} {11}},\ \bibinfo {pages} {7028} (\bibinfo {year} {2021})}\BibitemShut {NoStop}%
\bibitem [{\citenamefont {Noonan}\ and\ \citenamefont {Lambiotte}(2021)}]{noonan2021dynamics}%
  \BibitemOpen
  \bibfield  {author} {\bibinfo {author} {\bibfnamefont {J.}~\bibnamefont {Noonan}}\ and\ \bibinfo {author} {\bibfnamefont {R.}~\bibnamefont {Lambiotte}},\ }\bibfield  {title} {\bibinfo {title} {Dynamics of majority rule on hypergraphs},\ }\href@noop {} {\bibfield  {journal} {\bibinfo  {journal} {Physical Review E}\ }\textbf {\bibinfo {volume} {104}},\ \bibinfo {pages} {024316} (\bibinfo {year} {2021})}\BibitemShut {NoStop}%
\bibitem [{\citenamefont {Kim}\ \emph {et~al.}(2024)\citenamefont {Kim}, \citenamefont {Lee}, \citenamefont {Min}, \citenamefont {Porter}, \citenamefont {Miguel},\ and\ \citenamefont {Goh}}]{goh2024}%
  \BibitemOpen
  \bibfield  {author} {\bibinfo {author} {\bibfnamefont {J.}~\bibnamefont {Kim}}, \bibinfo {author} {\bibfnamefont {D.-S.}\ \bibnamefont {Lee}}, \bibinfo {author} {\bibfnamefont {B.}~\bibnamefont {Min}}, \bibinfo {author} {\bibfnamefont {M.~A.}\ \bibnamefont {Porter}}, \bibinfo {author} {\bibfnamefont {M.~S.}\ \bibnamefont {Miguel}},\ and\ \bibinfo {author} {\bibfnamefont {K.-I.}\ \bibnamefont {Goh}},\ }\bibfield  {title} {\bibinfo {title} {Competition between group interactions and nonlinearity in voter dynamics on hypergraphs},\ }\href@noop {} {\bibfield  {journal} {\bibinfo  {journal} {arXiv preprint arXiv:2407.11261}\ } (\bibinfo {year} {2024})}\BibitemShut {NoStop}%
\bibitem [{\citenamefont {H{\'e}bert-Dufresne}\ \emph {et~al.}(2022)\citenamefont {H{\'e}bert-Dufresne}, \citenamefont {Waring}, \citenamefont {St-Onge}, \citenamefont {Niles}, \citenamefont {Kati~Corlew}, \citenamefont {Dube}, \citenamefont {Miller}, \citenamefont {Gotelli},\ and\ \citenamefont {McGill}}]{hebert2022source}%
  \BibitemOpen
  \bibfield  {author} {\bibinfo {author} {\bibfnamefont {L.}~\bibnamefont {H{\'e}bert-Dufresne}}, \bibinfo {author} {\bibfnamefont {T.~M.}\ \bibnamefont {Waring}}, \bibinfo {author} {\bibfnamefont {G.}~\bibnamefont {St-Onge}}, \bibinfo {author} {\bibfnamefont {M.~T.}\ \bibnamefont {Niles}}, \bibinfo {author} {\bibfnamefont {L.}~\bibnamefont {Kati~Corlew}}, \bibinfo {author} {\bibfnamefont {M.~P.}\ \bibnamefont {Dube}}, \bibinfo {author} {\bibfnamefont {S.~J.}\ \bibnamefont {Miller}}, \bibinfo {author} {\bibfnamefont {N.~J.}\ \bibnamefont {Gotelli}},\ and\ \bibinfo {author} {\bibfnamefont {B.~J.}\ \bibnamefont {McGill}},\ }\bibfield  {title} {\bibinfo {title} {Source-sink behavioural dynamics limit institutional evolution in a group-structured society},\ }\href@noop {} {\bibfield  {journal} {\bibinfo  {journal} {Royal Society Open Science}\ }\textbf {\bibinfo {volume} {9}},\ \bibinfo {pages} {211743} (\bibinfo {year} {2022})}\BibitemShut {NoStop}%
\bibitem [{\citenamefont {St-Onge}\ \emph {et~al.}(2024{\natexlab{a}})\citenamefont {St-Onge}, \citenamefont {H{\'e}bert-Dufresne},\ and\ \citenamefont {Allard}}]{st2024nonlinear}%
  \BibitemOpen
  \bibfield  {author} {\bibinfo {author} {\bibfnamefont {G.}~\bibnamefont {St-Onge}}, \bibinfo {author} {\bibfnamefont {L.}~\bibnamefont {H{\'e}bert-Dufresne}},\ and\ \bibinfo {author} {\bibfnamefont {A.}~\bibnamefont {Allard}},\ }\bibfield  {title} {\bibinfo {title} {Nonlinear bias toward complex contagion in uncertain transmission settings},\ }\href@noop {} {\bibfield  {journal} {\bibinfo  {journal} {Proceedings of the National Academy of Sciences of the United States of America}\ }\textbf {\bibinfo {volume} {121}},\ \bibinfo {pages} {e2312202121} (\bibinfo {year} {2024}{\natexlab{a}})}\BibitemShut {NoStop}%
\bibitem [{\citenamefont {St-Onge}\ \emph {et~al.}(2024{\natexlab{b}})\citenamefont {St-Onge}, \citenamefont {Burgio}, \citenamefont {Rosenblatt}, \citenamefont {Waring},\ and\ \citenamefont {H{\'e}bert-Dufresne}}]{st2024paradoxes}%
  \BibitemOpen
  \bibfield  {author} {\bibinfo {author} {\bibfnamefont {J.}~\bibnamefont {St-Onge}}, \bibinfo {author} {\bibfnamefont {G.}~\bibnamefont {Burgio}}, \bibinfo {author} {\bibfnamefont {S.~F.}\ \bibnamefont {Rosenblatt}}, \bibinfo {author} {\bibfnamefont {T.~M.}\ \bibnamefont {Waring}},\ and\ \bibinfo {author} {\bibfnamefont {L.}~\bibnamefont {H{\'e}bert-Dufresne}},\ }\bibfield  {title} {\bibinfo {title} {Paradoxes in the coevolution of contagions and institutions},\ }\href@noop {} {\bibfield  {journal} {\bibinfo  {journal} {Proceedings of the Royal Society B: Biological Sciences}\ }\textbf {\bibinfo {volume} {291}},\ \bibinfo {pages} {20241117} (\bibinfo {year} {2024}{\natexlab{b}})}\BibitemShut {NoStop}%
\bibitem [{\citenamefont {Mill{\'a}n}\ \emph {et~al.}(2025)\citenamefont {Mill{\'a}n}, \citenamefont {Sun}, \citenamefont {Giambagli}, \citenamefont {Muolo}, \citenamefont {Carletti}, \citenamefont {Torres}, \citenamefont {Radicchi}, \citenamefont {Kurths},\ and\ \citenamefont {Bianconi}}]{millan2025}%
  \BibitemOpen
  \bibfield  {author} {\bibinfo {author} {\bibfnamefont {A.~P.}\ \bibnamefont {Mill{\'a}n}}, \bibinfo {author} {\bibfnamefont {H.}~\bibnamefont {Sun}}, \bibinfo {author} {\bibfnamefont {L.}~\bibnamefont {Giambagli}}, \bibinfo {author} {\bibfnamefont {R.}~\bibnamefont {Muolo}}, \bibinfo {author} {\bibfnamefont {T.}~\bibnamefont {Carletti}}, \bibinfo {author} {\bibfnamefont {J.~J.}\ \bibnamefont {Torres}}, \bibinfo {author} {\bibfnamefont {F.}~\bibnamefont {Radicchi}}, \bibinfo {author} {\bibfnamefont {J.}~\bibnamefont {Kurths}},\ and\ \bibinfo {author} {\bibfnamefont {G.}~\bibnamefont {Bianconi}},\ }\bibfield  {title} {\bibinfo {title} {Topology shapes dynamics of higher-order networks},\ }\href@noop {} {\bibfield  {journal} {\bibinfo  {journal} {Nature Physics}\ } (\bibinfo {year} {2025})},\ \bibinfo {note} {available at \url{https://doi.org/10.1038/s41567-024-02757-w}}\BibitemShut {NoStop}%
\bibitem [{\citenamefont {Cisneros-Velarde}\ and\ \citenamefont {Bullo}(2022)}]{cisneros2021multi}%
  \BibitemOpen
  \bibfield  {author} {\bibinfo {author} {\bibfnamefont {P.}~\bibnamefont {Cisneros-Velarde}}\ and\ \bibinfo {author} {\bibfnamefont {F.}~\bibnamefont {Bullo}},\ }\bibfield  {title} {\bibinfo {title} {Multi-group {SIS} epidemics with simplicial and higher-order interactions},\ }\href@noop {} {\bibfield  {journal} {\bibinfo  {journal} {IEEE Transactions on Control of Network Systems}\ }\textbf {\bibinfo {volume} {9}},\ \bibinfo {pages} {695} (\bibinfo {year} {2022})}\BibitemShut {NoStop}%
\bibitem [{\citenamefont {Landry}\ and\ \citenamefont {Restrepo}(2020)}]{landry2020effect}%
  \BibitemOpen
  \bibfield  {author} {\bibinfo {author} {\bibfnamefont {N.~W.}\ \bibnamefont {Landry}}\ and\ \bibinfo {author} {\bibfnamefont {J.~G.}\ \bibnamefont {Restrepo}},\ }\bibfield  {title} {\bibinfo {title} {The effect of heterogeneity on hypergraph contagion models},\ }\href@noop {} {\bibfield  {journal} {\bibinfo  {journal} {Chaos: An Interdisciplinary Journal of Nonlinear Science}\ }\textbf {\bibinfo {volume} {30}},\ \bibinfo {pages} {103117} (\bibinfo {year} {2020})}\BibitemShut {NoStop}%
\bibitem [{\citenamefont {Skardal}\ and\ \citenamefont {Arenas}(2020)}]{skardal2020higher}%
  \BibitemOpen
  \bibfield  {author} {\bibinfo {author} {\bibfnamefont {P.~S.}\ \bibnamefont {Skardal}}\ and\ \bibinfo {author} {\bibfnamefont {A.}~\bibnamefont {Arenas}},\ }\bibfield  {title} {\bibinfo {title} {Higher order interactions in complex networks of phase oscillators promote abrupt synchronization switching},\ }\href@noop {} {\bibfield  {journal} {\bibinfo  {journal} {Communications Physics}\ }\textbf {\bibinfo {volume} {3}},\ \bibinfo {pages} {218} (\bibinfo {year} {2020})}\BibitemShut {NoStop}%
\bibitem [{\citenamefont {Adhikari}\ \emph {et~al.}(2023)\citenamefont {Adhikari}, \citenamefont {Restrepo},\ and\ \citenamefont {Skardal}}]{adhikari2023synchronization}%
  \BibitemOpen
  \bibfield  {author} {\bibinfo {author} {\bibfnamefont {S.}~\bibnamefont {Adhikari}}, \bibinfo {author} {\bibfnamefont {J.~G.}\ \bibnamefont {Restrepo}},\ and\ \bibinfo {author} {\bibfnamefont {P.~S.}\ \bibnamefont {Skardal}},\ }\bibfield  {title} {\bibinfo {title} {Synchronization of phase oscillators on complex hypergraphs},\ }\href@noop {} {\bibfield  {journal} {\bibinfo  {journal} {Chaos: An Interdisciplinary Journal of Nonlinear Science}\ }\textbf {\bibinfo {volume} {33}},\ \bibinfo {pages} {033116} (\bibinfo {year} {2023})}\BibitemShut {NoStop}%
\bibitem [{\citenamefont {Prentice}\ and\ \citenamefont {Miller}(1996)}]{prentice1996pluralistic}%
  \BibitemOpen
  \bibfield  {author} {\bibinfo {author} {\bibfnamefont {D.~A.}\ \bibnamefont {Prentice}}\ and\ \bibinfo {author} {\bibfnamefont {D.~T.}\ \bibnamefont {Miller}},\ }\bibfield  {title} {\bibinfo {title} {Pluralistic ignorance and the perpetuation of social norms by unwitting actors},\ }\href@noop {} {\bibfield  {journal} {\bibinfo  {journal} {Advances in Experimental Social Psychology}\ }\textbf {\bibinfo {volume} {28}},\ \bibinfo {pages} {161} (\bibinfo {year} {1996})}\BibitemShut {NoStop}%
\bibitem [{\citenamefont {Miller}\ and\ \citenamefont {McFarland}(1987)}]{miller1987pluralistic}%
  \BibitemOpen
  \bibfield  {author} {\bibinfo {author} {\bibfnamefont {D.~T.}\ \bibnamefont {Miller}}\ and\ \bibinfo {author} {\bibfnamefont {C.}~\bibnamefont {McFarland}},\ }\bibfield  {title} {\bibinfo {title} {Pluralistic ignorance: When similarity is interpreted as dissimilarity},\ }\href@noop {} {\bibfield  {journal} {\bibinfo  {journal} {Journal of Personality and Social Psychology}\ }\textbf {\bibinfo {volume} {53}},\ \bibinfo {pages} {298} (\bibinfo {year} {1987})}\BibitemShut {NoStop}%
\bibitem [{\citenamefont {Janis}(1971)}]{janis2008groupthink}%
  \BibitemOpen
  \bibfield  {author} {\bibinfo {author} {\bibfnamefont {I.~L.}\ \bibnamefont {Janis}},\ }\bibfield  {title} {\bibinfo {title} {Groupthink},\ }\href@noop {} {\bibfield  {journal} {\bibinfo  {journal} {Psychology Today}\ }\textbf {\bibinfo {volume} {5}},\ \bibinfo {pages} {43} (\bibinfo {year} {1971})}\BibitemShut {NoStop}%
\bibitem [{\citenamefont {Kureh}\ and\ \citenamefont {Porter}(2020)}]{kureh2020}%
  \BibitemOpen
  \bibfield  {author} {\bibinfo {author} {\bibfnamefont {Y.~H.}\ \bibnamefont {Kureh}}\ and\ \bibinfo {author} {\bibfnamefont {M.~A.}\ \bibnamefont {Porter}},\ }\bibfield  {title} {\bibinfo {title} {Fitting in and breaking up: {A} nonlinear version of coevolving voter models},\ }\href@noop {} {\bibfield  {journal} {\bibinfo  {journal} {Physical Review E}\ }\textbf {\bibinfo {volume} {101}},\ \bibinfo {pages} {062303} (\bibinfo {year} {2020})}\BibitemShut {NoStop}%
\bibitem [{\citenamefont {Bicchieri}\ and\ \citenamefont {Mercier}(2014)}]{bicchieri2014norms}%
  \BibitemOpen
  \bibfield  {author} {\bibinfo {author} {\bibfnamefont {C.}~\bibnamefont {Bicchieri}}\ and\ \bibinfo {author} {\bibfnamefont {H.}~\bibnamefont {Mercier}},\ }\bibfield  {title} {\bibinfo {title} {Norms and beliefs: {H}ow change occurs},\ }in\ \href@noop {} {\emph {\bibinfo {booktitle} {The Complexity of Social Norms}}}\ (\bibinfo  {publisher} {Springer-Verlag},\ \bibinfo {address} {Heidelberg, Germany},\ \bibinfo {year} {2014})\ pp.\ \bibinfo {pages} {37--54}\BibitemShut {NoStop}%
\bibitem [{Note1()}]{Note1}%
  \BibitemOpen
  \bibinfo {note} {One can absorb the quantities $\langle k \rangle $ and $\langle q \rangle $ in Eqs.~\protect \eqref {xbar} and \protect \eqref {ybar} into the model parameters $a$, $b$, $c$, and $d$.}\BibitemShut {Stop}%
\bibitem [{\citenamefont {Nikolay}\ and\ \citenamefont {Svetoslav}(2015)}]{sigmoid2015}%
  \BibitemOpen
  \bibfield  {author} {\bibinfo {author} {\bibfnamefont {K.}~\bibnamefont {Nikolay}}\ and\ \bibinfo {author} {\bibfnamefont {M.}~\bibnamefont {Svetoslav}},\ }\href@noop {} {\emph {\bibinfo {title} {Sigmoid Functions: Some Approximation and Modelling Aspects}}}\ (\bibinfo  {publisher} {Lambert Academic Publishing},\ \bibinfo {address} {London, UK},\ \bibinfo {year} {2015})\BibitemShut {NoStop}%
\bibitem [{\citenamefont {Bizyaeva}\ \emph {et~al.}(2023)\citenamefont {Bizyaeva}, \citenamefont {Franci},\ and\ \citenamefont {Leonard}}]{bizyaeva2022nonlinear}%
  \BibitemOpen
  \bibfield  {author} {\bibinfo {author} {\bibfnamefont {A.}~\bibnamefont {Bizyaeva}}, \bibinfo {author} {\bibfnamefont {A.}~\bibnamefont {Franci}},\ and\ \bibinfo {author} {\bibfnamefont {N.~E.}\ \bibnamefont {Leonard}},\ }\bibfield  {title} {\bibinfo {title} {Nonlinear opinion dynamics with tunable sensitivity},\ }\href@noop {} {\bibfield  {journal} {\bibinfo  {journal} {IEEE Transactions on Automatic Control}\ }\textbf {\bibinfo {volume} {68}},\ \bibinfo {pages} {1415} (\bibinfo {year} {2023})}\BibitemShut {NoStop}%
\bibitem [{\citenamefont {Newman}(2018)}]{newman2018networks}%
  \BibitemOpen
  \bibfield  {author} {\bibinfo {author} {\bibfnamefont {M.~E.~J.}\ \bibnamefont {Newman}},\ }\href@noop {} {\emph {\bibinfo {title} {Networks}}},\ \bibinfo {edition} {2nd}\ ed.\ (\bibinfo  {publisher} {Oxford University Press},\ \bibinfo {address} {Oxford, UK},\ \bibinfo {year} {2018})\BibitemShut {NoStop}%
\bibitem [{\citenamefont {Kiss}\ \emph {et~al.}(2017)\citenamefont {Kiss}, \citenamefont {Miller},\ and\ \citenamefont {Simon}}]{kiss2017}%
  \BibitemOpen
  \bibfield  {author} {\bibinfo {author} {\bibfnamefont {I.~Z.}\ \bibnamefont {Kiss}}, \bibinfo {author} {\bibfnamefont {J.~C.}\ \bibnamefont {Miller}},\ and\ \bibinfo {author} {\bibfnamefont {P.~L.}\ \bibnamefont {Simon}},\ }\href@noop {} {\emph {\bibinfo {title} {Mathematics of Epidemics on Networks: {F}rom Exact to Approximate Models}}}\ (\bibinfo  {publisher} {Springer International Publishing},\ \bibinfo {address} {Cham, Switzerland},\ \bibinfo {year} {2017})\BibitemShut {NoStop}%
\bibitem [{\citenamefont {de~Arruda}\ \emph {et~al.}(2024)\citenamefont {de~Arruda}, \citenamefont {Aleta},\ and\ \citenamefont {Moreno}}]{ferraz2024}%
  \BibitemOpen
  \bibfield  {author} {\bibinfo {author} {\bibfnamefont {G.~F.}\ \bibnamefont {de~Arruda}}, \bibinfo {author} {\bibfnamefont {A.}~\bibnamefont {Aleta}},\ and\ \bibinfo {author} {\bibfnamefont {Y.}~\bibnamefont {Moreno}},\ }\bibfield  {title} {\bibinfo {title} {Contagion dynamics on higher-order networks},\ }\href@noop {} {\bibfield  {journal} {\bibinfo  {journal} {Nature Reviews Physics}\ }\textbf {\bibinfo {volume} {6}},\ \bibinfo {pages} {468} (\bibinfo {year} {2024})}\BibitemShut {NoStop}%
\bibitem [{\citenamefont {Chodrow}(2020)}]{chodrow2020configuration}%
  \BibitemOpen
  \bibfield  {author} {\bibinfo {author} {\bibfnamefont {P.~S.}\ \bibnamefont {Chodrow}},\ }\bibfield  {title} {\bibinfo {title} {Configuration models of random hypergraphs},\ }\href@noop {} {\bibfield  {journal} {\bibinfo  {journal} {Journal of Complex Networks}\ }\textbf {\bibinfo {volume} {8}},\ \bibinfo {pages} {cnaa018} (\bibinfo {year} {2020})}\BibitemShut {NoStop}%
\bibitem [{\citenamefont {Melnik}\ \emph {et~al.}(2014)\citenamefont {Melnik}, \citenamefont {Porter}, \citenamefont {Mucha},\ and\ \citenamefont {Gleeson}}]{melnik2014}%
  \BibitemOpen
  \bibfield  {author} {\bibinfo {author} {\bibfnamefont {S.}~\bibnamefont {Melnik}}, \bibinfo {author} {\bibfnamefont {M.~A.}\ \bibnamefont {Porter}}, \bibinfo {author} {\bibfnamefont {P.~J.}\ \bibnamefont {Mucha}},\ and\ \bibinfo {author} {\bibfnamefont {J.~P.}\ \bibnamefont {Gleeson}},\ }\bibfield  {title} {\bibinfo {title} {Dynamics on modular networks with heterogeneous correlations},\ }\href@noop {} {\bibfield  {journal} {\bibinfo  {journal} {Chaos: An Interdisciplinary Journal of Nonlinear Science}\ }\textbf {\bibinfo {volume} {24}},\ \bibinfo {pages} {023106} (\bibinfo {year} {2014})}\BibitemShut {NoStop}%
\bibitem [{\citenamefont {Barrio}\ \emph {et~al.}(2020)\citenamefont {Barrio}, \citenamefont {Coombes}, \citenamefont {Desroches}, \citenamefont {Fenton}, \citenamefont {Luther},\ and\ \citenamefont {Pueyo}}]{barrio2020excitable}%
  \BibitemOpen
  \bibfield  {author} {\bibinfo {author} {\bibfnamefont {R.}~\bibnamefont {Barrio}}, \bibinfo {author} {\bibfnamefont {S.}~\bibnamefont {Coombes}}, \bibinfo {author} {\bibfnamefont {M.}~\bibnamefont {Desroches}}, \bibinfo {author} {\bibfnamefont {F.}~\bibnamefont {Fenton}}, \bibinfo {author} {\bibfnamefont {S.}~\bibnamefont {Luther}},\ and\ \bibinfo {author} {\bibfnamefont {E.}~\bibnamefont {Pueyo}},\ }\bibfield  {title} {\bibinfo {title} {Excitable dynamics in neural and cardiac systems},\ }\href@noop {} {\bibfield  {journal} {\bibinfo  {journal} {Communications in Nonlinear Science and Numerical Simulation}\ }\textbf {\bibinfo {volume} {86}},\ \bibinfo {pages} {105275} (\bibinfo {year} {2020})}\BibitemShut {NoStop}%
\bibitem [{\citenamefont {D'Errico}(2014)}]{matlabpackage}%
  \BibitemOpen
  \bibfield  {author} {\bibinfo {author} {\bibfnamefont {J.}~\bibnamefont {D'Errico}},\ }\bibfield  {title} {\bibinfo {title} {{Adaptive Robust Numerical Differentiation}}} (\bibinfo {year} {2014}),\ \bibinfo {note} {version 1.6, available at \url{https://www.mathworks.com/matlabcentral/fileexchange/13490-adaptive-robust-numerical-differentiation} (last accessed 22 February 2025)}\BibitemShut {NoStop}%
\bibitem [{\citenamefont {Porter}\ and\ \citenamefont {Gleeson}(2016)}]{porter2016}%
  \BibitemOpen
  \bibfield  {author} {\bibinfo {author} {\bibfnamefont {M.~A.}\ \bibnamefont {Porter}}\ and\ \bibinfo {author} {\bibfnamefont {J.~P.}\ \bibnamefont {Gleeson}},\ }\href@noop {} {\emph {\bibinfo {title} {Dynamical Systems on Networks: {A} Tutorial}}},\ Vol.~\bibinfo {volume} {4}\ (\bibinfo  {publisher} {Springer International Publishing},\ \bibinfo {address} {Cham, Switzerland},\ \bibinfo {year} {2016})\ \bibinfo {note} {{Frontiers in Applied Dynamical Systems: Reviews and Tutorials}}\BibitemShut {NoStop}%
\bibitem [{\citenamefont {Landry}\ and\ \citenamefont {Restrepo}(2022)}]{landry2022hypergraph}%
  \BibitemOpen
  \bibfield  {author} {\bibinfo {author} {\bibfnamefont {N.~W.}\ \bibnamefont {Landry}}\ and\ \bibinfo {author} {\bibfnamefont {J.~G.}\ \bibnamefont {Restrepo}},\ }\bibfield  {title} {\bibinfo {title} {Hypergraph assortativity: A dynamical systems perspective},\ }\href@noop {} {\bibfield  {journal} {\bibinfo  {journal} {Chaos: An Interdisciplinary Journal of Nonlinear Science}\ }\textbf {\bibinfo {volume} {32}} (\bibinfo {year} {2022})}\BibitemShut {NoStop}%
\bibitem [{\citenamefont {Landry}\ and\ \citenamefont {Restrepo}(2023)}]{landry2023opinion}%
  \BibitemOpen
  \bibfield  {author} {\bibinfo {author} {\bibfnamefont {N.~W.}\ \bibnamefont {Landry}}\ and\ \bibinfo {author} {\bibfnamefont {J.~G.}\ \bibnamefont {Restrepo}},\ }\bibfield  {title} {\bibinfo {title} {Opinion disparity in hypergraphs with community structure},\ }\href@noop {} {\bibfield  {journal} {\bibinfo  {journal} {Physical Review E}\ }\textbf {\bibinfo {volume} {108}},\ \bibinfo {pages} {034311} (\bibinfo {year} {2023})}\BibitemShut {NoStop}%
\bibitem [{\citenamefont {M{\"a}s}(2019)}]{mas2019}%
  \BibitemOpen
  \bibfield  {author} {\bibinfo {author} {\bibfnamefont {M.}~\bibnamefont {M{\"a}s}},\ }\bibfield  {title} {\bibinfo {title} {Challenges to simulation validation in the social sciences. {A} critical rationalist perspective},\ }in\ \href@noop {} {\emph {\bibinfo {booktitle} {Computer Simulation Validation: Fundamental Concepts, Methodological Frameworks, and Philosophical Perspectives}}},\ \bibinfo {editor} {edited by\ \bibinfo {editor} {\bibfnamefont {C.}~\bibnamefont {Beisbart}}\ and\ \bibinfo {editor} {\bibfnamefont {N.~J.}\ \bibnamefont {Saam}}}\ (\bibinfo  {publisher} {Springer International Publishing},\ \bibinfo {address} {Cham, Switzerland},\ \bibinfo {year} {2019})\ pp.\ \bibinfo {pages} {857--879}\BibitemShut {NoStop}%
\bibitem [{\citenamefont {Bak-Coleman}\ \emph {et~al.}(2021)\citenamefont {Bak-Coleman}, \citenamefont {Alfano}, \citenamefont {Barfuss}, \citenamefont {Bergstrom}, \citenamefont {Centeno}, \citenamefont {Couzin}, \citenamefont {Donges}, \citenamefont {Galesic}, \citenamefont {Gersick}, \citenamefont {Jacquet}, \citenamefont {Kao}, \citenamefont {Moran}, \citenamefont {Romanczuk}, \citenamefont {Rubenstein}, \citenamefont {Tombak}, \citenamefont {Van~Bavel},\ and\ \citenamefont {Weber}}]{bak2021}%
  \BibitemOpen
  \bibfield  {author} {\bibinfo {author} {\bibfnamefont {J.~B.}\ \bibnamefont {Bak-Coleman}}, \bibinfo {author} {\bibfnamefont {M.}~\bibnamefont {Alfano}}, \bibinfo {author} {\bibfnamefont {W.}~\bibnamefont {Barfuss}}, \bibinfo {author} {\bibfnamefont {C.~T.}\ \bibnamefont {Bergstrom}}, \bibinfo {author} {\bibfnamefont {M.~A.}\ \bibnamefont {Centeno}}, \bibinfo {author} {\bibfnamefont {I.~D.}\ \bibnamefont {Couzin}}, \bibinfo {author} {\bibfnamefont {J.~F.}\ \bibnamefont {Donges}}, \bibinfo {author} {\bibfnamefont {M.}~\bibnamefont {Galesic}}, \bibinfo {author} {\bibfnamefont {A.~S.}\ \bibnamefont {Gersick}}, \bibinfo {author} {\bibfnamefont {J.}~\bibnamefont {Jacquet}}, \bibinfo {author} {\bibfnamefont {A.~B.}\ \bibnamefont {Kao}}, \bibinfo {author} {\bibfnamefont {R.~E.}\ \bibnamefont {Moran}}, \bibinfo {author} {\bibfnamefont {P.}~\bibnamefont {Romanczuk}}, \bibinfo {author} {\bibfnamefont {D.~I.}\ \bibnamefont {Rubenstein}}, \bibinfo {author} {\bibfnamefont {K.~J.}\ \bibnamefont {Tombak}}, \bibinfo
  {author} {\bibfnamefont {J.~J.}\ \bibnamefont {Van~Bavel}},\ and\ \bibinfo {author} {\bibfnamefont {E.~U.}\ \bibnamefont {Weber}},\ }\bibfield  {title} {\bibinfo {title} {Stewardship of global collective behavior},\ }\href@noop {} {\bibfield  {journal} {\bibinfo  {journal} {Proc. Natl. Acad. Sci. USA}\ }\textbf {\bibinfo {volume} {118}},\ \bibinfo {eid} {e2025764118} (\bibinfo {year} {2021})}\BibitemShut {NoStop}%
\bibitem [{\citenamefont {Iacopini}\ \emph {et~al.}(2024)\citenamefont {Iacopini}, \citenamefont {Karsai},\ and\ \citenamefont {Barrat}}]{iacopini2024}%
  \BibitemOpen
  \bibfield  {author} {\bibinfo {author} {\bibfnamefont {I.}~\bibnamefont {Iacopini}}, \bibinfo {author} {\bibfnamefont {M.}~\bibnamefont {Karsai}},\ and\ \bibinfo {author} {\bibfnamefont {A.}~\bibnamefont {Barrat}},\ }\bibfield  {title} {\bibinfo {title} {The temporal dynamics of group interactions in higher-order social networks},\ }\href@noop {} {\bibfield  {journal} {\bibinfo  {journal} {Nature Communications}\ }\textbf {\bibinfo {volume} {15}},\ \bibinfo {pages} {7391} (\bibinfo {year} {2024})}\BibitemShut {NoStop}%
\bibitem [{\citenamefont {Torres}\ \emph {et~al.}(2021)\citenamefont {Torres}, \citenamefont {Blevins}, \citenamefont {Bassett},\ and\ \citenamefont {Eliassi-Rad}}]{torres2021}%
  \BibitemOpen
  \bibfield  {author} {\bibinfo {author} {\bibfnamefont {L.}~\bibnamefont {Torres}}, \bibinfo {author} {\bibfnamefont {A.~S.}\ \bibnamefont {Blevins}}, \bibinfo {author} {\bibfnamefont {D.}~\bibnamefont {Bassett}},\ and\ \bibinfo {author} {\bibfnamefont {T.}~\bibnamefont {Eliassi-Rad}},\ }\bibfield  {title} {\bibinfo {title} {The why, how, and when of representations for complex systems},\ }\href@noop {} {\bibfield  {journal} {\bibinfo  {journal} {SIAM Review}\ }\textbf {\bibinfo {volume} {63}},\ \bibinfo {pages} {435} (\bibinfo {year} {2021})}\BibitemShut {NoStop}%
\bibitem [{\citenamefont {Fosdick}\ \emph {et~al.}(2018)\citenamefont {Fosdick}, \citenamefont {Larremore}, \citenamefont {Nishimura},\ and\ \citenamefont {Ugander}}]{fosdick2018configuring}%
  \BibitemOpen
  \bibfield  {author} {\bibinfo {author} {\bibfnamefont {B.~K.}\ \bibnamefont {Fosdick}}, \bibinfo {author} {\bibfnamefont {D.~B.}\ \bibnamefont {Larremore}}, \bibinfo {author} {\bibfnamefont {J.}~\bibnamefont {Nishimura}},\ and\ \bibinfo {author} {\bibfnamefont {J.}~\bibnamefont {Ugander}},\ }\bibfield  {title} {\bibinfo {title} {Configuring random graph models with fixed degree sequences},\ }\href@noop {} {\bibfield  {journal} {\bibinfo  {journal} {Siam Review}\ }\textbf {\bibinfo {volume} {60}},\ \bibinfo {pages} {315} (\bibinfo {year} {2018})}\BibitemShut {NoStop}%
\bibitem [{\citenamefont {Newman}(2009)}]{newman2009}%
  \BibitemOpen
  \bibfield  {author} {\bibinfo {author} {\bibfnamefont {M.~E.~J.}\ \bibnamefont {Newman}},\ }\bibfield  {title} {\bibinfo {title} {Random graphs with clustering},\ }\href@noop {} {\bibfield  {journal} {\bibinfo  {journal} {Physical Review Letters}\ }\textbf {\bibinfo {volume} {103}},\ \bibinfo {pages} {058701} (\bibinfo {year} {2009})}\BibitemShut {NoStop}%
\bibitem [{\citenamefont {Miller}(2009)}]{miller2009}%
  \BibitemOpen
  \bibfield  {author} {\bibinfo {author} {\bibfnamefont {J.~C.}\ \bibnamefont {Miller}},\ }\bibfield  {title} {\bibinfo {title} {Percolation and epidemics in random clustered networks},\ }\href@noop {} {\bibfield  {journal} {\bibinfo  {journal} {Physical Review E}\ }\textbf {\bibinfo {volume} {80}},\ \bibinfo {pages} {020901} (\bibinfo {year} {2009})}\BibitemShut {NoStop}%
\end{thebibliography}%


\end{document}